\documentclass[amsmath,amssymb,nofootinbib,showpacs,twocolumn,nofootinbib]{revtex4-1}
\usepackage{dcolumn}
\usepackage{bm}
\usepackage{graphicx}
\usepackage{epstopdf}
\usepackage{epsfig}
\usepackage{color}
\usepackage{wasysym}
\usepackage{natbib}
\usepackage{twoopt}
\usepackage{dashrule}
\definecolor{slateblue}{rgb}{0.2,0.22,0.6}
\usepackage[breaklinks=true,colorlinks=true,linkcolor=slateblue,citecolor=slateblue,urlcolor=slateblue,pdfauthor={JieFeng},pdftitle={THMMCMC}]{hyperref}
\definecolor{ColorTitle}{cmyk}{0,.88,.77,.40}

\newcommand{\AMS}{AMS-02}
\newcommand{\etal}{et al.}
\newcommand{\eg}{\textit{e.g.}} 
\newcommand{\ie}{\textit{i.e.}} 
\newcommand{\Dragon}{\texttt{DRAGON}}
\newcommand{\CRMC}{\texttt{CRMC}}
\newcommand{\captionsize}{\footnotesize}

\newcommand{\R}{\ensuremath{\mathcal{R}}}
\renewcommand{\P}{\ensuremath{\mathcal{P}}} 

\newcommand{\p}{\ensuremath{p}}
\newcommand{\n}{\ensuremath{n}}
\newcommand{\pp}{\ensuremath{p}-\ensuremath{p}}
\newcommand{\pC}{\ensuremath{p}-\ensuremath{C}}

\newcommand{\He}{He}
\newcommand{\Li}{Li}
\newcommand{\Be}{Be}
\newcommand{\B}{B}
\newcommand{\C}{C}
\newcommand{\N}{N}
\newcommand{\F}{F}
\newcommand{\Oxy}{O}
\newcommand{\Al}{Al}
\newcommand{\Fe}{Fe}

\newcommand{\BC}{{B}/{C}}
\newcommand{\LiC}{{Li/C}}
\newcommand{\BeB}{{Be/B}}

\newcommand{\HeHe}{\ensuremath{{}^{3}}{He}/\ensuremath{{}^{4}}{He}}
\newcommand{\dHe}{\ensuremath{{}^{2}}{H}/\ensuremath{{}^{4}}{He}}
\newcommand{\BeBe}{\ensuremath{{}^{10}}{Be}/\ensuremath{{}^{9}}{Be}}
\newcommand{\BeTen}{\ensuremath{{}^{10}}Be}
\newcommand{\BTen}{\ensuremath{{}^{10}}B}

\newcommand{\eplus}{\ensuremath{e^{+}}}

\newcommand{\epfrac}{e\ensuremath{^{+}}/(e\ensuremath{^{-}}\,+\,e\ensuremath{^{+}})} 

\newcommand{\pbarp}{\textsf{\ensuremath{\bar{p}/p}}}
\newcommand{\pbar}{\textsf{\ensuremath{\bar{p}}}}
\newcommand{\nbar}{\textsf{\ensuremath{\bar{n}}}}

\begin{document}
\title{Bayesian analysis of spatial-dependent cosmic-ray propagation:\\ astrophysical background of antiprotons and positrons}

\author{Jie Feng\,$^{1,2}$}
\email{jie.feng@cern.ch}
\author{Nicola Tomassetti\,$^{3,4}$}
\email{nicola.tomassetti@cern.ch}
\author{Alberto Oliva\,$^{5}$}
\email{alberto.oliva@cern.ch}
\affiliation{$^{1}$School of Physics, Sun Yat-Sen University, Guangzhou 510275, China;}
\affiliation{$^{2}$Institute of Physics, Academia Sinica, Nankang, Taipei 11529, Taiwan;}
\affiliation{$^{3}$Universit{\`a} degli Studi di Perugia \& INFN-Perugia, I-06100 Perugia, Italy;}
\affiliation{$^{4}$LPSC, Universit\'e Grenoble-Alpes, CNRS/IN2P3, F-38026 Grenoble, France,}
\affiliation{$^{5}$Centro de Investigaciones Energ\'{e}ticas, Medioambientales Tecnol\'{o}gicas, E-28040 Madrid, Spain}
\begin{abstract}
  The \AMS{} experiment has reported a new measurement of the antiproton/proton ratio in Galactic cosmic rays (CRs).
  In the energy range $E\sim$\,60--450\,GeV, this ratio is found to be remarkably constant.
  Using recent data on CR proton, helium, carbon fluxes, $^{10}$Be/$^{9}$Be and B/C ratios, we have performed a global Bayesian
  analysis based on a Markov Chain Monte-Carlo sampling algorithm under a ``two halo model'' of CR propagation. 
  In this model, CRs are allowed to experience a different type of diffusion when they propagate in the region close of the Galactic disk.
  We found that the vertical extent of this region is about 900 pc above and below the disk, and the corresponding diffusion 
  coefficient scales with energy as $D\propto\,E^{0.15}$, describing well the observations on primary CR spectra, 
  secondary/primary ratios and anisotropy. Under this model we have carried out improved calculations of antiparticle spectra arising 
  from secondary CR  production and their corresponding uncertainties.
  We made use of Monte-Carlo generators and accelerator data to assess the antiproton production cross-sections and their uncertainties.
  While the positron excess requires the contribution of additional unknown sources, we found that the new \AMS{} antiproton data are consistent, 
  within the estimated uncertainties, with our calculations based on secondary production.
\end{abstract}
\maketitle

\section{Introduction}    
\label{Sec::Introduction} 

An increase in the accuracy of cosmic-ray (CR) spectra and composition measurements is driving us 
to a deeper understanding of the fundamental physics processes that CRs experience in the Galaxy. 
The current era appears particularly promising for CR physics. New-generation detection experiments 
such as the Payload for Antimatter Matter Exploration and Light-nuclei Astrophysics (PAMELA) and 
the Alpha Magnetic Spectrometer (\AMS) in space, or the Advanced Thin Ionization Calorimeter (ATIC-2) and 
the Cosmic Ray Energetics and Mass (CREAM) on balloon have brought important results in the physics of CR 
propagation \citep{CR:Review}. 
At the same time, valuable pieces of information are being achieved by the large wealth of $\gamma$-ray data coming from 
space or ground-based telescopes such as Fermi Large Area Telescope (Fermi-LAT) or High Energy Stereoscopic System (H.E.S.S.). 
Other space missions recently launched (\ie, the Dark Matter Particle Explorer DAMPE and the CALorimetric Electron Telescope CALET)
or awaiting launch (CREAM for the International Space Station, ISS-CREAM)
will soon provide high-quality CR data in the multi-TeV energy scale \cite{Maestro:2015ICRC}. 
In this ``golden age''  of CR physics measurements, several unexpected features are being discovered in the CR energy spectrum, 
hence the study of CR propagation has become more important than ever \citep{Serpico:2015ICRC}.
We can mention the unexpected \emph{increase} of the positron fraction \epfrac{} at $E\sim$\,10--300\,GeV
\cite{AMS02:Leptons,FermiLAT:2011ab,PAMELA:Results}, 
or the puzzling spectral hardening in proton and helium at $E\gtrsim$\,200 GeV/nucleon
\cite{PAMELA:Results,Yoon:2011aa,Panov:2011ak, Aguilar:2015PHe}, 
both established by a series of measurements operated on balloons and in space. 
On top of this, the \AMS{} collaboration has now released new precision data on the antiproton-to-proton (\pbarp) ratio 
between $\sim$\,0.5 and 450\,GeV of kinetic energy reporting that, above $\sim$\,60\,GeV, 
the ratio remains remarkably \emph{constant} with energy \cite{Aguilar:2016pbar}.

In standard models of CR propagation, antimatter particles are only produced by collisions of high-energy 
nuclei with the gas of the interstellar medium (ISM), from which the \pbarp{} ratio or the positron fraction 
are expected to \emph{decrease} with energy, very roughly, as fast as the boron-to-carbon (\BC) ratio does. 
To interpret the positron excess, it seems to be unavoidable the introduction of extra sources of high-energy positrons such as 
dark matter particle annihilation \cite{Leptons:DM} or $e^{\pm}$ pair production mechanisms inside nearby pulsars \cite{Leptons:Pulsar}. 
A corresponding unaccountable excess in CR antiprotons would considerably point toward the DM annihilation scenario. 
The new \AMS{} data on the \pbarp{} ratio are in fact at tension with standard-model predictions 
of secondary antiproton production, and this tension generated widespread interest \cite{Pbar:Sec}. 
Interpretations of these data in terms of TeV scale dark matter have been suggested \cite{Pbar:TeVDM}.
However, the level of \emph{astrophysical background} in CR antiprotons is far from being understood, because conventional 
models of CR propagation suffers from large uncertainties and intrinsic limitations in describing the fine structures of new 
observations \citep{Giesen:2015ufa,Pbar:Unc}. 
For instance, the high-energy spectral hardening observed in proton and helium cannot be explained within the traditional picture 
based on linear diffusive-shock-acceleration (DSA) followed by homogeneous propagation in the ISM \cite{Serpico:2015ICRC}. 
On the other hands, these features are offering a clue for making substantial advances in understanding the physics of Galactic CRs.
Several recent works suggested the need of accounting for nearby source contributions to the observed 
flux of CRs \citep{Kachelriess2015SNR,TomassettiDonato,Tomassetti2015TwoSnr},
non-linear effects in their propagation \citep{Blasi:2012yr,Recchia:2016,ThoudamHorandel:2014}, or spatial-dependent
diffusion processes \cite{Erlykin&Gaggero,Johannesson:2016,Kappl:2016}.
We remark that understanding these effects is of crucial importance for assessing the astrophysical antimatter background.

In this paper, we report the results of a large scan on the CR injection and propagation parameter space
in order to provide a robust prediction for the secondary production of antiprotons and positrons along with their uncertainties. 
To describe the CR transport in the Galaxy, we set up a numerical implementation of a two-halo (THM) scenario of diffusive
propagation \cite{TomassettiSH,Guo&Jin}, where CRs are allowed to experience a
different type of diffusion when they propagate closer to the Galactic plane. 
To assess the uncertainties on the CR acceleration and transport parameters, we adopt a Markov Chain Monte-Carlo (MCMC) sampling technique
with the incorporation of a large set of nuclear data such as proton, helium, and carbon fluxes,
the \BC{} elemental ratio, and the \BeBe{} isotopic ratio.
The MCMC-based determination of the key model parameters and their probability density functions, accounting for
correlations between the free parameters, allows to determine well defined uncertainty bounds for the secondary antimatter flux
calculations which constitute the astrophysical background for the search of new-physics signals.
We will also review systematic uncertainties in the model arising from the solar modulation effect and from antiparticle production cross-sections.
We found that the new $\pbarp$ ratio measured by \AMS{} is fairly well described by a THM propagation model within the estimated 
level of uncertainties, while the excess of $\eplus$ requires the presence of extra sources. 

This paper is organized as follows. In Sect.\,\ref{Sec::Propagation_model} we outline our CR propagation calculations and physics analysis setup.
In particular we describe the numerical implementation of our model and the methodology adopted for the determination of the key parameters.
In Sect.\,\ref{Sec::Results} we present our results, we discuss the probability distribution inferred on the parameters and their inter-dependence 
in connection with the CR physics observables. We also use our best-fit model to calculate the astrophysical background of CR antiprotons and positrons.
We discuss our finding and some critical aspects of our calculations in Sect.\,\ref{Sec::Discussion}.
We conclude with Sect.\,\ref{Sec::Conclusions} by summarizing the main focus points of this work. 
In Appendix~\ref{Sec::ApppendixB}, we provide additional information on the antiproton production cross-sections and their uncertainties.

\section{Calculations}          
\label{Sec::Propagation_model}  

\subsection{Cosmic ray propagation modeling} 
\label{Sec::Propagation}                     

The CR energy spectrum from sub-GeV to multi-TeV energy arises from a combination of DSA
and diffusive propagation processes occurring in our Galaxy. CRs are believed to originate in Galactic sources such as
supernova remnants (SNRs). They are injected into the ISM after being DSA accelerated to rigidity power-law
spectra $Q\propto \R^{-\nu}$, where $\R=pc/Ze$, with index $\nu\sim$\,2--2.4.
Their diffusive propagation in the interstellar turbulence is usually described by means of a rigidity dependent 
diffusion coefficient $D(\R)\propto \R^{\delta}$, with $\delta\sim$\,0.2--0.7.
The propagation region is usually modeled as an extended cylinder with radius $R_{C}$ of $\sim$\,20\,kpc and half-height $L$ of a few kpc.
The propagation of all CR species is often described by a two-dimensional transport equation with boundary conditions at $r=r_{\rm max}$ and $z=\pm L$:
\begin{equation}\label{Eq::DiffusionTransport}
  \frac{\partial {\psi}}{\partial t} = Q + \vec{\nabla}\cdot (D\vec{\nabla}{\psi}) - {{\psi}}{\Gamma} + \frac{\partial}{\partial E} (b\, {\psi})  \,,
\end{equation}
where $\psi=\psi(E,r,z)$ is the particle number density as a function of energy and space coordinates,
$\Gamma= \beta c n \sigma$ is the destruction rate for collisions off gas nuclei, with density $n$, at velocity $\beta c$ 
and cross-section $\sigma$. The source term $Q$ is split into a primary term, $Q_{\rm pri}$, 
and a secondary production term $Q_{\rm sec}= \sum_{\rm j} \Gamma_{j}^{\rm sp} \psi_{\rm j}$, from spallation of 
heavier $j$--type nuclei with rate $\Gamma_{j}^{\rm sp}$. 
The term $b(E)=\--\frac{dE}{dt}$ describes ionization and Coulomb losses, as well as radiative cooling of CR leptons.
The steady-state solution of Eq.\,\ref{Eq::DiffusionTransport} for the CR flux near the solar system
reflects the combined effects of injection and propagation. Approximately, primary CR components such as
protons, \He, \C, \Oxy, or \Fe, with $Q\approx Q_{\rm pri}$, have power-law spectra  $\psi_{p} \sim Q/D \propto \R^{-\delta-\nu}$.
The equilibrium spectra  $\psi_{s}$ of purely secondary CRs such as \Li-\Be-\B{} nuclei or antiparticles, with $Q = Q_{\rm sec}$, 
are $R^{\delta}$-times steeper than those of primary CRs, so that $\psi_{s}/\psi_{p}\sim \R^{-\delta}$.
Similar expectations are given for antiparticle/particle ratios such the positron fraction \epfrac{} or the \pbarp{} ratio.
Another key property is the size of the propagation halo, which can be probed by measuring radioactive secondary CR nuclei $^{10}$\Be{} or $^{26}$\Al.
To assess the halo height is of fundamental importance in dark matter searches. 
Propagation models may also include other effects such as reacceleration, usually described as a diffusion in momentum space,
or convective transport induced by Galactic wind. While these effects have been successful in reproducing the shape of the \BC{} ratio, 
they generate problems for primary elements such as \p{} and \He{} in the $\sim$\,0.1-10\,GeV/n energy region. 
Since we are focusing on high-energy region, and no consensus has been reached on reacceleration or convection, we disregard such effects.
It is also important to stress, however, that the traditional picture as illustrated above
is unable to account for the high-energy spectral hardening recently observed in CR protons and nuclei.
Possible approaches to model these features involve \citep{Vladimirov:2012,Serpico:2015ICRC}: 
(i) introduction of multi-component populations for the CR flux;
(ii) revisitation of CR injection, $Q_{\rm pri}(\R)$, reflecting non-linear and/or time-dependent DSA; or 
(iii) modification of CR diffusion, $D(\R)$, accounting for nonlinear effects in  CR propagation or spatial-dependent diffusion.
Here we adopt the latter scenario which is well supported by recent works and, as we will show,
it leads in general to conservative predictions for the production of secondary antiparticles in the ISM.

\subsection{Numerical implementation} 
\label{Sec::NumericalImplementation}  

We set up a spatial dependent scenario of CR propagation in two halos,
which is the simplest physically consistent generalization of the standard models
that are able to account for the recent observations of CR hadrons.
It consists in allowing CRs to experience a different (shallower) type of diffusion when 
they propagate in the proximity of the Galactic disk.
In practice, 
this idea is implemented by splitting the cylindrical propagation region into two $z$-symmetric halos
characterized by different diffusion properties: the inner halo, which surrounds the disk for a few hundred pc,
and the outer halo, an extended regions of a few kpc which surrounds the inner halo.
Numerically, our model is implemented under the \Dragon{} code of CR propagation, which is well suited for handling
CR diffusion in inhomogeneous media \cite{Evoli:2008}. 
We introduced a modification of the finite-differencing scheme in the solver \citep{Guo&Jin,TomassettiSH},
in order to obtain a spatial-dependent and non-separable diffusion coefficient $D=D(z,\R)$. 
To test CR diffusion close to the Galactic disk, we set up a non-equidistant spatial grid where the pitch
from two consecutive nodes increases with the coordinate $|z|$. We adopt a diffusion coefficient of the following form:
\begin{equation}\label{Eq::DiffusionCoefficient}
  D(\R,z) =  
  \left\{
  \begin{array}{r@{\; \;}l}
    D_{0} \beta^{\eta}{\left(\frac{\R}{\R_0}\right)}^{\delta}  \quad\quad\quad&  (|z| < \xi L) \\
    \chi D_0 \beta^{\eta}{\left(\frac{\R}{\R_0}\right)}^{\delta +\Delta} &  (|z| > \xi L) 
  \end{array}
  \right.
\end{equation}
where a connecting function of the type $F(z)= \left(z/L\right)^{n}$ is used to 
ensure a smooth transition of the parameters $\chi$ and $\Delta$ across the two zones \citep{Guo&Jin}.
The parameter $D_{0}$ sets the normalization of the diffusion in the disk 
at the reference rigidity $\R_{0}\equiv 0.25$\,GV, while $\chi D_{0}$ is used for the outer halo. 
The low-energy diffusion is shaped by the factor $\beta^{\eta}$, where $\beta=v/c$ is the particle velocity
divided by the speed of light and $\eta$ is set to be $-0.4$ \cite{Evoli:2008}. 
The parameter $\delta$ is the diffusion scaling index in the inner halo (with $|z|< \xi L$) 
while $\delta+\Delta$ is that of the outer halo ($\xi L < |z|<L$), and $L$ is the half-height of the whole diffusion region.

This scenario of CR propagation is supported by radio observation on other galaxies such as NGC 891, NGC 253 or M51 \citep{Heesen:2009}
and favored by the observed level of CR anisotropy at multi-TeV energies \citep{Guo&Jin}. 
Interpretations at the origin of the two zones have been proposed in terms of different types of Galactic turbulence, \eg,
SNR-driven and CR-driven turbulence
that are supported by $\gamma$--ray observations on latitudinal and radial dependence of CR spectra \citep{Erlykin&Gaggero}.
It was recently argued that two diffusion regimes may be connected with advective CR tansport on 
self-induced Galactic wind \cite{Recchia:2016}. 
In this work we use the data to constraint the relevant parameters of CR transport in the two regions, 
namely, $D_0$, $\chi$, $\delta$, $\Delta$, $L$, and $\xi$.
In addition to the six transport parameters, we introduce two parameters describing injection:
the spectral index $\nu$ of proton injection, and the spectral index
difference $\Delta\nu$ between protons and all other primary nuclei such as \He, \C, \Oxy, \Fe.
The latter parameter accounts for the recently observed spectral difference between proton and \He{} \citep{Aguilar:2015PHe,PAMELA:Results}, 
being the former $\Delta\nu$-times steeper than the latter. Since no spectral differences have been
observed on heavier nuclei, we adopt the same slope for all $Z>1$ primary spectra.
With the use of non-universal injection indices we are ascribing the origin of the observed \p/\He{} anomaly to
an intrinsic DSA acceleration mechanism, as proposed recently \citep{OhiraMalkov}. 
Note however that there are explanations for the \p/\He{} anomaly which do not require composition-dependent 
acceleration mechanisms \citep{Tomassetti2015TwoSnr,ThoudamHorandel:2014,Vladimirov:2012}.

\subsection{Parameter sampling and data sets} 
\label{Sec::MCMC}                             

\begin{table}[!htbp]
  \label{Tab::Priors}%
  \begin{tabular}{ c c c c c}
    \hline\hline
    \textbf{parameter} & $\,$\textbf{units}$\,$ & $\,$\textbf{prior}$\,$ & $\,$\textbf{minimum}$\,$ & $\,$\textbf{maximum}$\,$
    \tabularnewline
    \hline
    \hline
    $ L $ & kpc & 6.8 & 2.5 & 9.5 \tabularnewline
    \hline
    $ D_{0}$ & $10^{28}$\,{cm}$^{2}$\,{s}$^{-1}$  & 1.7 & 0.5 & 5.0 \tabularnewline
    \hline
    $\delta$ & \dots & 0.16 & 0. & 0.6 \tabularnewline
    \hline
    $\Delta$ & \dots & 0.56 & 0.2 & 1.2 \tabularnewline
    \hline
    $\xi$ &\dots & 0.14 & 0.08 & 0.6 \tabularnewline
    \hline
    $\chi$ & \dots & 0.35 & 0.2 & 1.2 \tabularnewline
    \hline
    $\Delta\nu$ &\dots & 0.09 & 0.03 & 0.15 \tabularnewline
    \hline
    $\nu$ & \dots & 2.27 & 2.0 & 2.6 \tabularnewline
    \hline\hline
  \end{tabular}
  \caption{\captionsize%
    Prior values and ranges for the injection and transport parameters.}
\end{table}

Our scan operates in a eight-dimensional parameter space. 
To perform an efficient sampling, we make use of the MCMC method based on the Bayesian inference.
Recent works demonstated that the MCMC method is a practical and powerful tool for CR propagation physics 
analysis \cite{PutzeMaurin:2010,Trotta:2011,Liu:2012,Korsmeier2016,Johannesson:2016}.
Bayesian inference is about the quantification and propagation of uncertainties, 
expressed in terms of probability, in light of observations of the system.
Our specific goal is to estimate the probability density functions (PDFs) of our set of free parameters for 
the following inputs: 
(i) an underlying model of CR propagation which provides the link between physics observables and parameters: 
(ii) a defined sets of experimental data, and  
(iii) the prior distributions of the input parameters.
The output PDFs are given as posterior probabilities which quantify the change in the degree of belief 
one can have in the model parameters in the light of the data.
Our parameter set is given by the vector ${\bm {\theta}}= \{ D_0, \chi, \delta, \Delta, L, \xi, \gamma, \Delta\gamma \}$, 
while the available observations are the ensemble of the data vector $\textbf{D}$. 
From the Bayes theorem, the posterior distribution reads: 
\begin{equation}\label{Eq::Bayes}
  \P({\bm{\theta}}|\textbf{D}) = \frac{\P(\textbf{D}|{\bm{\theta}}) \P({\bm{\theta}})}{\P(\textbf{D})} \,.
\end{equation}
The posterior probability $\P({\bm{\theta}}|\textbf{D})$ depends on the likelihood function
$\mathcal{L}({\bm{\theta}}) = \P(\textbf{D}|{\bm{\theta}})$ and on the prior probability distribution $\P({\bm{\theta}})$.
The latter expresses our state of knowledge about about the parameters to be inferred.

We employ the Metropolis -- Hastings  algorithm, which makes use of the Monte-Carlo method to generate
large sequences of random samples \cite{Metropolis:1953am}. 
In practice the chains are built according to a arbitrary proposal distribution, $q\left(\bm\theta_n,\bm\theta_{n+1}\right)$,
which is used to MC-generate a 
new paramter configuration $\bm\theta_{n+1}$ starting from $\bm\theta_{n}$. 
The new configuration $\bm\theta_{n+1}$ is then randomly accepted (or rejected) according to a probability $\alpha$ (or $1-\alpha$),
which is built as:
\begin{equation}\label{Eq::MCMCprobability}
  \alpha\left(\bm\theta_n,\bm\theta_{n+1}\right) =  \text{min}\left\{1, 
  \frac{\P\left(\bm\theta_{n+1}\right)q\left(\bm\theta_{n+1},\bm\theta_n\right)}{\P\left(\bm\theta_n\right)q\left(\bm\theta_n,\bm\theta_{n+1}\right)} 
  \right\}\,,
\end{equation}
where $\P\left(\bm\theta\right)$ is the distribution of $\bm\theta$ expected from the sample. 
If $\bm\theta_{n+1}$ is rejected, the new state is again $\bm\theta_{n}$.
The transition probability is then 
$T\left(\bm\theta_n, \bm\theta_{n+1}\right)= \alpha\left(\bm\theta_n,\bm\theta_{n+1}\right)q\left(\bm\theta_n,\bm\theta_{n+1}\right)$.
The target distribution converges to 
$\P\left(\bm\theta_{n+1}\right)T\left(\bm\theta_{n+1}, \bm\theta_n\right)=\P\left(\bm\theta_n\right)T\left(\bm\theta_n, \bm\theta_{n+1}\right)$.
After a large sampling, we eventually get $\P\left({\bm\theta}\right)$ as the equilibrium distribution for the chain.
The likelihood is defined as $\mathcal{L}({\bm\theta})= \exp{ \left( -\frac{1}{2}\chi^{2}\right) }$, 
where the $\chi^{2}({\bm\theta})$ function is built from the data and model output as:
\begin{equation}\label{Eq::ChiSquare}
  \chi^{2}({\bm\theta}) = \sum_{k=1}^{N_{D}} \left( \frac{ y_{k}^{\rm exp} - y_{k}^{\rm th}({\bm\theta})}{\sigma_{k}} \right)^{2} 
\end{equation}
This equation establishes a link between model parameters ${\bm\theta}$
and experimental data $y^{\rm exp}$ with corresponding uncertainties $\sigma_{k}$.
The injection and transport parameter are summarized in Table\,\ref{Tab::Priors}, where
their initial values (priors) and the corresponding ranges are listed.
The observed quantities $y^{\rm exp}$ and those predicted $y^{\rm th}({\bm\theta})$ consist in CR fluxes or nuclear ratios.
In this work we use the most accurate and recent experimental CR data sets available.
Primary CR spectra are taken from the recent \AMS{} measurements of proton and \He{} fluxes \cite{Aguilar:2015PHe} up to TeV/n energies,
and from the CREAM experiments \cite{Yoon:2011aa} which covers the multi-TeV energy region. 
The carbon spectrum is constrained using the data from PAMELA and CREAM \citep{PAMELA:Results,Yoon:2011aa}.
The \BC{} ratio has been measured by several of space-based and balloon-borne experiments.
We use the data recently released from the PAMELA collaboration \cite{PAMELA:Results} 
along with measurements from AMS-01 \cite{Aguilar:2010hm}, ATIC-2 \cite{Panov:2007fe}, 
and TeV/n energy data from CREAM \cite{Ahn:2008my} and TRACER \cite{Obermeier:2012vg}.
We do not include older data, as  various studies pointed out the possibility that systematic uncertainties 
in old \BC{} data are considerably underestimated \citep{Trotta:2011,PutzeMaurin:2010}.
We include in our study several data on the \BeBe{} ratio \cite{BeBeData}. 
For this ratio, measurements are in quite scarce and affected by sizable uncertainties.
To account for the solar modulation effect we adopt the simple \emph{force-field} approximation, where the
strength of the modulation effect is expressed in terms of the solar modulation potential $\phi$ \cite{Potgieter2013}. 
For a given $Z$-charged particles at given epoch in the course of the 11-year cycle, 
its kinetic energy is shifted via the relation $E^{\odot}=E^{\rm IS} - \frac{|Z|}{A} \phi$, while the modulated
CR density is given by:
\begin{equation}\label{Eq::ForceField}
  \psi^{\odot} = \frac{(E+ mc^{2})^{2}- m^{2}c^{4}}{(E+mc^{2} +|Z|e\phi)^{2}-m^{2}c^{4}} \times \psi^{\rm IS}(E + |Z|e\phi) \,.
\end{equation}
The value of the parameter $\phi$ depends on the solar activity and can be different for various data sets.
Its determination can performed simultaneously with the other parameters, in principle, by accounting for it the global likelihood. 
In principle, solar modulation could be even better modeled using three-dimensional simulations of particle transport in the heliosphere
which accounts for adiabatic losses, particle drift effects, or anisotropic diffusion \citep{Potgieter2013}. 
However these models have several free parameters and their application is not feasible with the available computing resources. 
Here we  adopt a simple approach based on the consideration that the intensity of the modulation effect 
decreases with energy (being $\lesssim\,$1\% level at energy above a few $\sim$\,10\,GeV/nucleon),
and it is suppressed for nuclear ratios such as \BC{} or \BeBe. 
For the differential spectra of \p, \He, and \C, we build our $\chi^{2}({\bm\theta})$ function using only data
above a minimal energy $E_{\rm min}=45$\,GeV/n. At these energies the effect of solar modulation is 
smaller than the experimental uncertainties of the data.
For the \BC{} ratio, similarly, we set $E_{\rm min}=2$\,GeV/n. At these energies the ratio is measured by the PAMELA experiment
during the 2009-2010 solar minimum, with a modulation potential of $\phi\approx$\,300\,MV \cite{PAMELA:Results,Ghelfi:2016pcv}. 
For the \BeBe{} ratio we set $E_{\rm min}=0.1$\,GeV/n. This ratio have been measured only below $\sim$\,5\,GeV/n
by several experiments, in the course of the last decades, characterized by different levels of solar activity.
For all data-sets we adopt multiple parameters $\phi$ and we account for their corresponding 
uncertainties $\delta\phi\cong\,50$\,MV \cite{Ghelfi:2016pcv}. 
Hence we estimate the impact of $\delta\phi$ in all observables in order to get residual uncertainties $\sigma_{\phi}$. 
For all the data and energy ranges considered, we account for the estimated uncertainties $\sigma_{\phi}$ 
in the construction of $\chi^{2}({\bm\theta})$, by adding $\sigma_{\phi}$ in quadrature to the
uncertainties of experimental data. 
With this procedure, the uncertainty in solar modulation will be inglobated in the MCMC posterior distributions.
Nonetheless, with the chosen data sets and energy ranges, solar modulation uncertainties are always smaller in 
comparison with the experimental errors.
We again discuss solar modulation in Sect.\,\ref{Sec::Results} for the calculations of \eplus{} and \pbar{} fluxes.

\section{Results and discussion}  
\label{Sec::Results}              

\subsection{Parameters and degeneracies} 
\label{Sec::ResultsParameters}           

\begin{figure*}[!t]
  \includegraphics[width=1.0\textwidth]{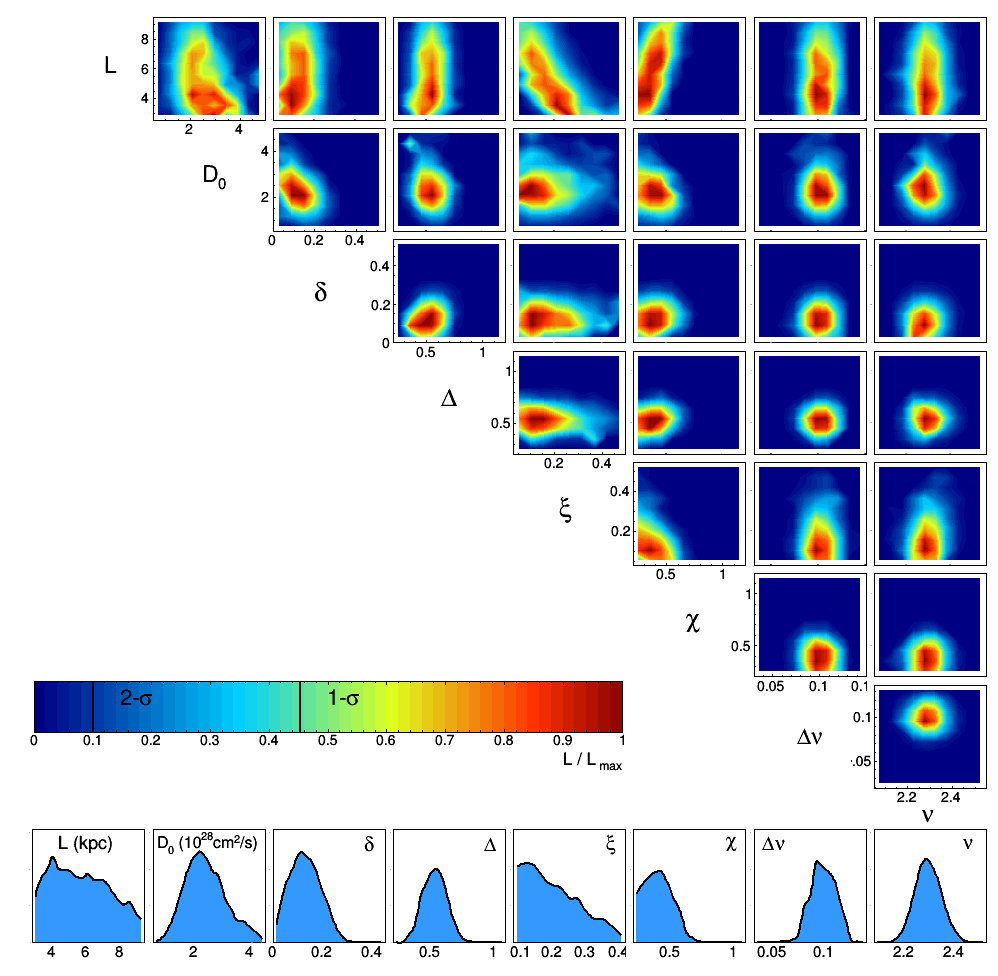} 
  \caption{\captionsize%
    Marginalized posterior distributions of all model parameters.
    The contour plots show the two-dimensional distributions of all pairs of parameter, 
    where the color scale represents the likelihood valaue normalized to the maximum likelihood. 
    The levels corresponding to  1-$\sigma$ and 2-$\sigma$ regions are indicated in the color bar.
    The bottom histograms represent the one-dimensional PDFs of all parameters.
  }
  \label{Fig::ccContour}
\end{figure*}

With the global MCMC-based parameter scan performed over the entire data set, we have obtained the posterior 
distributions for all parameters of the likelihood. In the triangular plot of Fig.\,\ref{Fig::ccContour} we show 
the two-dimensional distributions for all combinations of parameters in terms of contour plots.
Their marginalized posterior PDFs are shown in the bottom panels of the figures. 
The results are also summarized in Table\,\ref{Tab::Posteriors} where we list the best-fit values
of the parameters, corresponding to the maximum likelihood, along with their most probable values
or posterior modes and with their posterior means. The posterior mean of a parameter $\theta$ 
is computed as the expectation values over its PDF: 
\begin{equation}\label{Eq::PosteriorMean}
  \langle  {\theta} \rangle \equiv \int {\theta}\, \P({\theta}|\textbf{D})\, d{\theta} \,.
\end{equation}
The $1-\sigma$ and $2-\sigma$ confidence levels for all parameters, calculated from their marginalized PDFs, are also listed in the table. 
It can be seen that several parameters show well-behaved distributions. 
%
%
In particular the quantities $\Delta\nu$ and $\nu$ appear well constrained and essentially uncorrelated with other parameters. 
These parameters specify the acceleration sector and, with some dependence on the transport parameters, are sensitive primary 
CR data such as proton and helium. However, in the THM formulation, data on primary spectra may also allow to resolve transport 
parameters, such as the quantity $\Delta$ which is directly linked to the spectral index variation of all primary particles.
In general, best-fit values and most probable values do not coincide, due to the presence of degeneracy between parameters. 
Degeneracy, related to a lack of sensitivity of the likelihood in particular regions of the parameter space, 
gives also rise to parameter correlation. 
From Fig.\,\ref{Fig::ccContour}, it can be seen a clear anti-correlation between
the parameters $\xi$ and $L$ describing the sizes of the two propagation regions.
These parameters are also correlated with $\chi$ and $D_{0}$ which control the diffusion in the two zones.  
The one-dimensional distributions of $L$ and $\xi$ are considerably asymmetric,
and in particular, the PDF of the halo half-height $L$ is found to be rather broad.
For this parameter we found no clear convergence with well-defined uncertainty bounds.
This problem is better inspected in in Sect.\,\ref{Sec::ResultsBeNuclei}.
A known degeneracy is the one between $\delta$ and $D_{0}$. It can be seen that these 
parameters are slightly negatively correlated.
With model calculations associated with the best-fit parameters, we reproduce the observations very well.
From the fit results on the parameters, we have also drawn the 68\,\% and 95\,\% envelopes of several physics observables.
To better illustrate the parameter inter-dependencies and their connection with the data,
in the following sub-sections we present our results for the basic sets of observations:
primary nuclei, secondary/primary ratios, unstable isotopes, antiprotons, positrons, and anisotropy.
\begin{table*}[!t]
  \label{Tab::Posteriors}
  \begin{tabular}{c c c c c c c c c}
    \hline\hline
    \textbf{parameter} & \textbf{unit} & \textbf{best-fit} & $\>\>\>$\textbf{posterior mean} & $\>\>\>$\textbf{posterior mode} & $\>\>\>$\textbf{$1 \sigma$-low} & $\>\>\>$\textbf{$1 \sigma$-up} & $\>\>\>$\textbf{$2 \sigma$-low} & $\>\>\>$\textbf{$2 \sigma$-up} \tabularnewline
    \hline\hline
    $L$ & kpc  &  5.67 & 6.71 &  4.15 & \dots & \dots & \dots & \dots  \tabularnewline
    \hline
    $D_{0}$ & 10$^{28}$\,cm$^{2}$\,s$^{-1}$  & 1.92 & 1.83 & 2.20 & 0.75 & 4.99 & 0.50 & 4.99\tabularnewline
    \hline
    $\delta$ & \dots & 0.15 & 0.18 & 0.13&  0.004 & 0.312 & 0.003 &  0.522 \tabularnewline
    \hline
    $\Delta$ & \dots & 0.52 & 0.58 & 0.55 &  0.426 & 0.823 & 0.240 &  1.076 \tabularnewline
    \hline
    $\xi$ & \dots & 0.15 & 0.19 & 0.12 & 0.090 & 0.593 & 0.080 & 0.594 \tabularnewline
    \hline
    $\chi$ & \dots & 0.36 & 0.42  & 0.42 & 0.212 & 0.923 & 0.200 & 1.119 \tabularnewline
    \hline
    $\Delta\nu$ & \dots & 0.096 & 0.096 & 0.100& 0.064 & 0.132 & 0.030 & 0.145 \tabularnewline
      \hline
      $\nu$ & \dots & 2.27 & 2.30 & 2.28 & 2.14 & 2.47 & 2.09 & 2.55 \tabularnewline
      \hline\hline
  \end{tabular}
  \caption{\captionsize%
    Results of the MCMC scan for the transport and injection parameters in terms of best-fit values, posterior means, and posterior modes, 
    along with their bounds for $1-\sigma$ and $2-\sigma$ fiducial ranges.}
\end{table*}

\subsection{Primary nuclei}        
\label{Sec::ResultsPrimaryNuclei}  
\begin{figure}[!th]
  \includegraphics[width=0.46\textwidth]{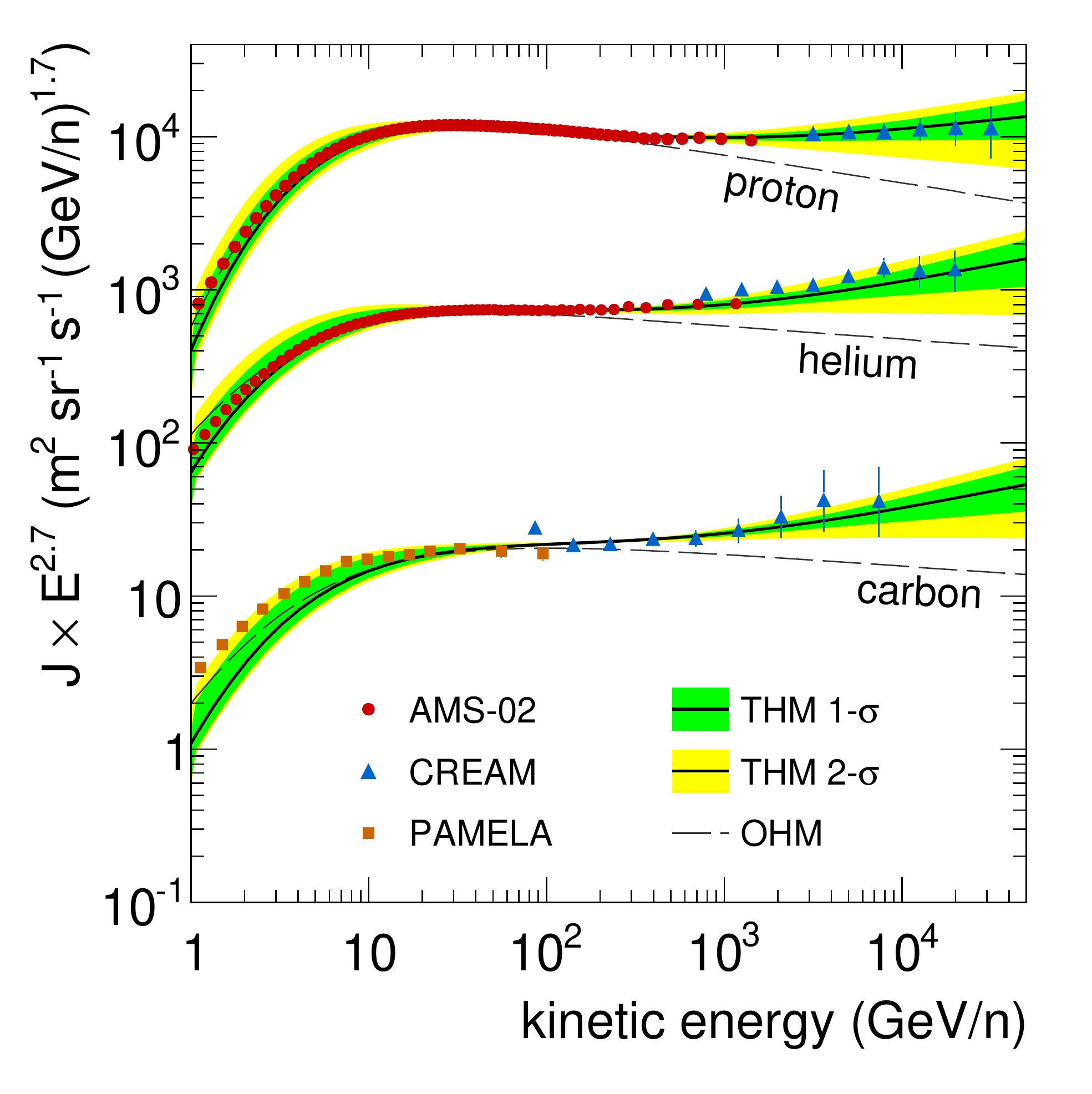} 
  \caption{\captionsize
    Model calculations and uncertainties for proton, \He\ and \C\ spectra compared with the experimental data 
    \cite{Aguilar:2015PHe, Yoon:2011aa, PAMELA:Results}.
  } 
  \label{Fig::ccProtonSpectrum}%
\end{figure}

\begin{figure}[!th]
  \includegraphics[width=0.46\textwidth]{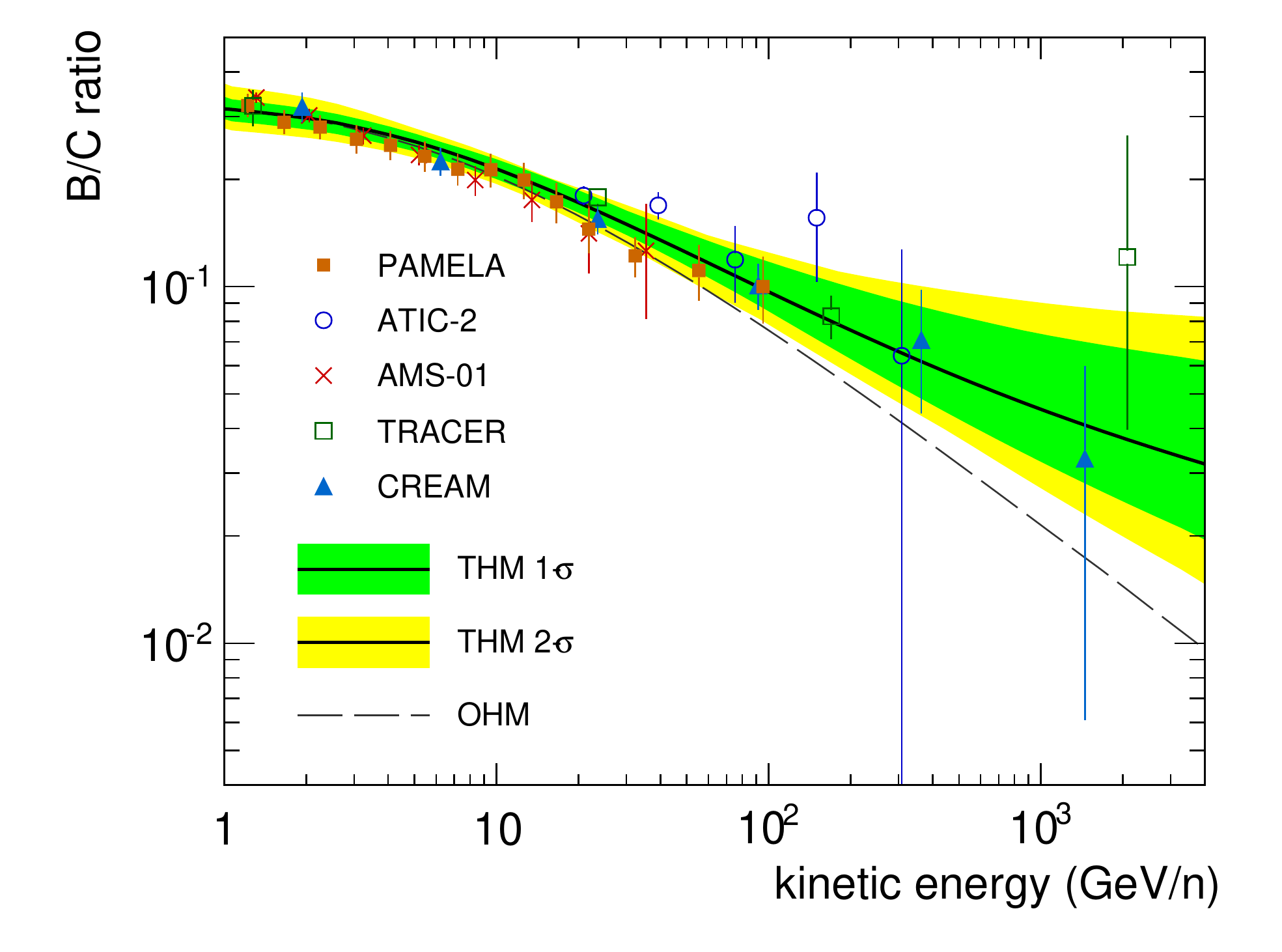}
  \includegraphics[width=0.46\textwidth]{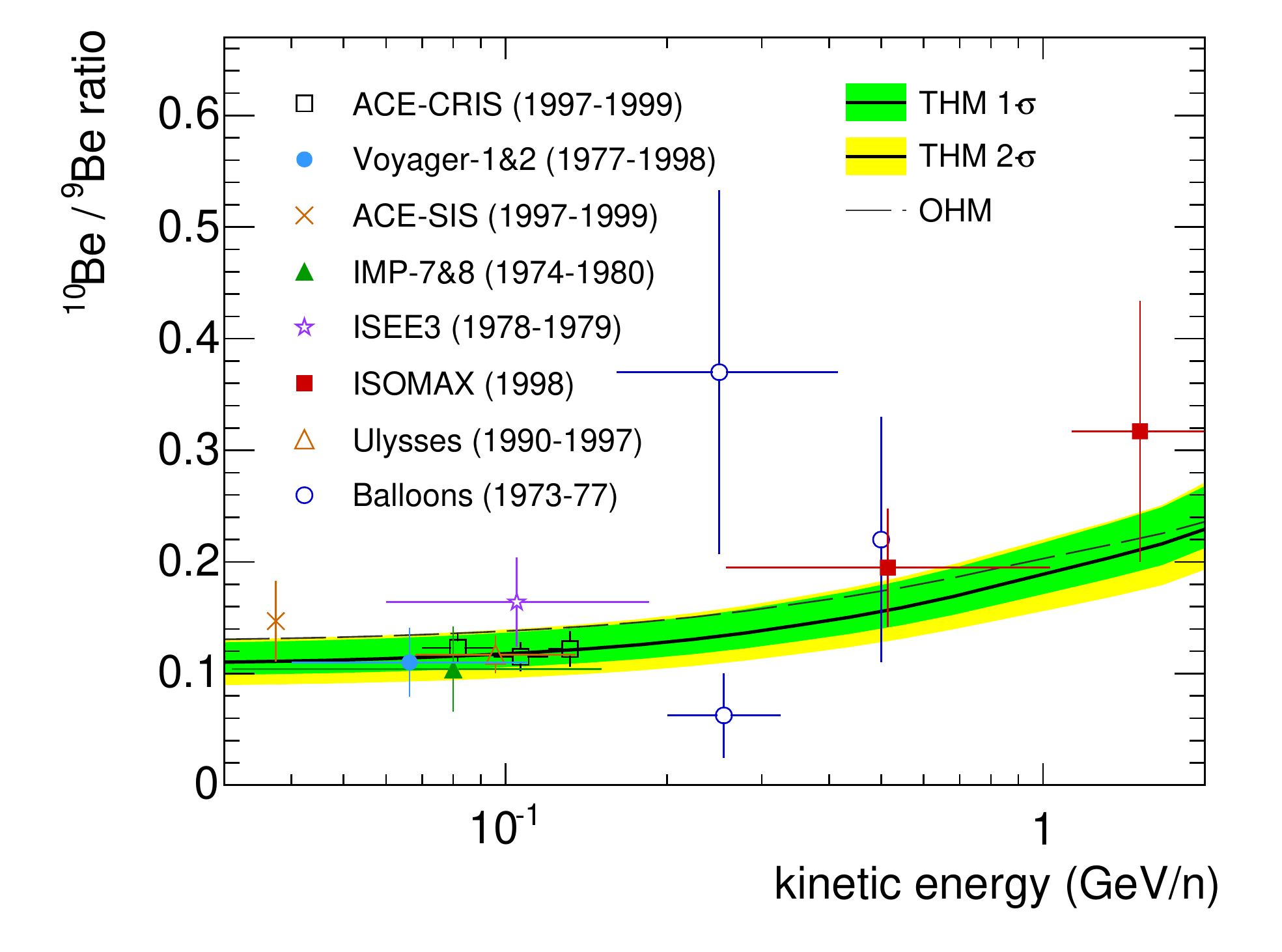}
  \caption{\captionsize%
    Top: model calculations and uncertainty band for the \BC\ ratio in comparison with the experimental
    data \cite{PAMELA:Results,Aguilar:2010hm,Panov:2007fe,Ahn:2008my,Obermeier:2012vg}. 
    Bottom: calculations for the \BeBe\ ratio in comparison with
    the measurements \cite{BeBeData}.
  } 
  \label{Fig::NucleiSpectra}
\end{figure}

The energy spectra of protons, helium, and carbon nuclei in CRs are shown in the top plot of Fig,\,\ref{Fig::ccProtonSpectrum}.
The THM calculations corresponding to the best-fit model are plotted as solid lines.
The shaded areas represent the 1-$\sigma$ (green) and 2-$\sigma$ (yellow) uncertainty bands. 
It can be seen that the model describes well the progressive hardening of the spectra at energies above $\sim$\,200\,GeV/n, in good accordance with the data. 
These results are consistent with the trends obtained from analytical calculations \cite{TomassettiSH}. 
In the figure, the thin dashed lines illustrate calculations from the standard ``one-halo'' diffusion model (OHM), 
that we show for comparison purpose. In this model we have imposed $\xi\equiv \chi \equiv$\,1.
The OHM parameter setting is determined using data at energies $E\lesssim$\,200 GeV/n. 
The parameter values are $L=4$\,kpc, $\delta=0.56$, and $D_{0}=0.52\times10^{28}\sim$\,{cm}$^{2}$\,{s}$^{-1}$ at $R_0\equiv 0.25$\,GV.
The most striking difference between the two models is apparent from the figure. An OHM scenario with standard power-law
injection spectra is unable to describe the data in the GeV-TeV energy range. 
The essential features of the THM can be understood under a purely diffusive regime (\ie, with interactions and energy losses neglected)
when passing in the 1D limit of $r_{\rm max}\ll L$. 
For an injection spectrum of the type $Q_{p}\sim\,\R^{-\nu}$, 
the primary flux at Earth, $J_{p}=\frac{c}{4\pi}\psi_{p}^{\odot}$,
can be approximately written in the form:
\begin{equation}\label{Eq::ApproximateTrendPrimary}
  J_{p} \sim \frac{L}{D_{0}}\left[\xi+\frac{1-\xi}{\chi} \left( \frac{\R}{\R_{0}} \right)^{-\Delta}\right] \left(\frac{\R}{\R_{0}}\right)^{-(\nu+\delta)}\,. 
\end{equation}
%
Equation\,\ref{Eq::ApproximateTrendPrimary} illustrates the degeneracy between transport and source parameters.
This behavior can be compared with the standard OHM expectations $J_{p} \sim (L/D_{0})\R^{-(\nu+\delta)}$
where the diffusion coefficient is a universal function of rigidity. 
The THM properties of the spectral hardening are controlled by the parameter $\chi$, which
sets the relative diffusion of the two zones, and by the parameter $\Delta$, which specifies the intensity of the upturn for all primary species. 
In practice, the first parameter sets the critical transition  rigidity (or energy) of the change in the spectral slope.
It is interesting to note that the parameter $\Delta$ appears well constrained. 
From the figure, one can see that the constraints provided by \AMS{} on \p{} and \He{} are very tight in the sub-TeV energy region.
The uncertainties at about 1\,GeV/nucleon energy include those coming from solar modulation, as discussed.
However, data multi-TeV energies are essential to determine $\Delta$, for which the best constraints are provided from the CREAM data.
This energy region is currently being investigated by several space experiments.
It can also be noted, from Eq.\,\ref{Eq::ApproximateTrendPrimary}, that the basic modification of
the CR injection spectra at relativistic rigidities are \emph{universal}, \ie, particle-independent, when expressed as function of rigidity.
Mass-dependent effects may appear in the low-energy region due to nuclear interactions
that are neglected in the above equation (but incorporated in our full THM implementation).
Since interactions increases with the nuclear mass number, one may expect the heavier nuclei show a
slightly harder flux in the low-energy region.
At high rigidities the resulting spectral change is expected to occur for all primary nuclei,
because it depends only on CR transport properties that have been determined by our global scan. 
For this reason, the model predictions for the carbon spectrum are mostly constrained by the more precise data on CR helium.
In fact the injection spectral indices are imposed to be the same for all $Z>1$ primary fluxes, 
while the subsequent propagation conditions are the same for all charged CR particles.
In contrast, for the calculation of the proton spectrum we have allowed for a different injection spectrum,
as recent measurements from CREAM, PAMELA and \AMS{} seem to suggest \citep{Vladimirov:2012,Aguilar:2015PHe}.
From our scan, the proton spectrum is found to be steeper by $\Delta\nu\approx\,0.1$ in terms of injection parameters.
Even though propagation effect may mitigate the spectral difference between the two species,  
the hypothesis of universal injection ($\Delta\nu=0$) is ruled out at 95\,\% of confidence level.
These results stand within the conception that the \p/\He{} anomaly is ascribed to intrinsic 
properties of accelerators \citep{Vladimirov:2012,OhiraMalkov}.
Besides, Eq.\,\ref{Eq::ApproximateTrendPrimary} illustrates clearly the degeneracies between parameters describing
transport and injection which involve combinations such as $\xi\,L/D_{0}$ or $\nu+\delta$. 
Similarly to standard diffusion models, complementary information from secondary CR nuclei is required to break these degeneracies.

\subsection{Secondary/primary ratios}   
\label{Sec::ResultsSecondaryNuclei}     

In fragmentation processes of relativistic CR nuclei with the ISM, the kinetic energy per nucleon
of secondary fragments ($s$) is approximately the same of that of their progenitor nuclei ($p$). 
Hence the for $p\rightarrow\,s$ fragmentation reactions, the ``source term'' of secondary nuclei
is approximately given by $Q_{s}\propto J_{p}$.
Thus the approximate THM behavior of secondary-to-primary ratios as function of rigidity reads:
\begin{equation}\label{Eq::ApproximateTrendSecondary}
  J_{s}/J_{p}\sim \frac{L}{D_{0}}\left[\xi+\frac{1-\xi}{\chi} \left( \frac{\R}{\R_{0}} \right)^{-\Delta}\right] \left(\frac{\R}{\R_{0}}\right)^{-\delta}\,. 
\end{equation}
It is interesting to note that the $R^{-\Delta}$ factor is present in both Eq.\,\ref{Eq::ApproximateTrendPrimary}
and Eq.\,\ref{Eq::ApproximateTrendSecondary}, \ie, the THM predicts the appearance of a spectral hardening in all secondary-to-primary ratios.
This feature is in general expected by any scenario where the spectral hardening is ascribed to CR propagation rather
than acceleration \citep{Vladimirov:2012,Blasi:2012yr}.
In Fig.\,\ref{Fig::NucleiSpectra} the \BC{} ratio calculations are shown in comparisons with the data.
From the THM calculations, the ratio has a tendency to flatten at kinetic energy above $\sim$\,100\,GeV/nucleon
which is driven by the data. 
This tendency results into a rather small best-fit value $\delta\approx$\,0.15
that we found for the diffusion spectral index in the inner halo.
Note that, in contrast to the parameter $\Delta$ which is directly constrained with primary CR spectra, 
the determination of $\delta$ requires information on secondary to primary ratios.
These results indicate that CR diffusion close to the Galactic disk has a rather weak rigidity dependence so that,
at high rigidities, CR propagation in the inner halo is significantly slower than that in the outer halo. 
It can be seen, however, that the TeV/nucleon energy region is affected by sizeable uncertainties due to the scarcity of \BC{} data.
The resulting PDF for the parameter $\delta$ is therefore quite broad.
Within 95\,\% of confidence level this parameter ranges from $\delta\sim\,0$ to $\delta\sim$\,0.5,
therefore encompassing its theoretically preferred values of $\delta=1/3$ (for a Kolmogorov type spectrum of
interstellar turbulence) and $\delta=1/2$ (for Iroshnikov-Kraichnan type diffusion).
These uncertainties are also related to unresolved parameter degeneracies.
For example the observed correlation between the scaling index $\delta$ and the diffusion
coefficient normalization $D_{0}$ arises from the limited constraining power of the \BC{} data at high energies. 
The determination of the key transport parameters, and more in general the understanding of CR diffusion at high energy, 
will be greatly improved with the data forthcoming \AMS{} and with those expected from DAMPE, CALET and ISS-CREAM
in the near future \citep{Maestro:2015ICRC}.
Complementary information to CR transport may come from isotopic secondary to primary ratios 
such as the \dHe{} and \HeHe{} ratios or from other secondary nuclei such as \Li, \Be, \F, or the sub-\Fe{} elements. 
In particular we stress that the lithium flux may be a powerful diagnostic tool for CR propagation models. 
Since \Li{} are produced from not only \C-\N-\Oxy{} collisions but also in so-called ``tertiary'' reactions processes, 
such as \B$\rightarrow$\Li{} and \Be$\rightarrow$\Li, which contribute appreciably in determining its spectral shape.
While the existing measurements of CR lithium are very scarce, \AMS{} will provide very precise \Li{} measurements
up to a few TV of rigidity. These data will be precious for testing astrophysical models of CR propagation.

\subsection{Unstable isotopes and halo height}       
\label{Sec::ResultsBeNuclei}\label{Sec::Disk_height} 

\begin{figure}[!t]
  \includegraphics[width=0.46\textwidth]{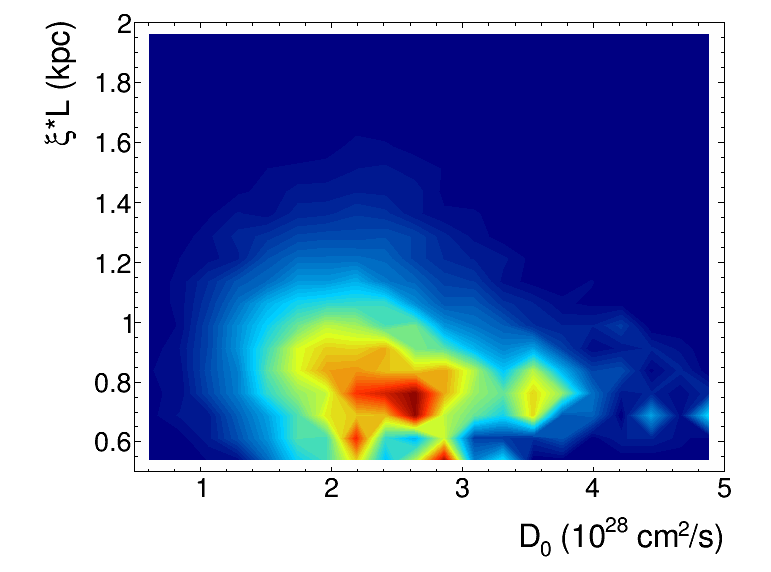}
  \includegraphics[width=0.46\textwidth]{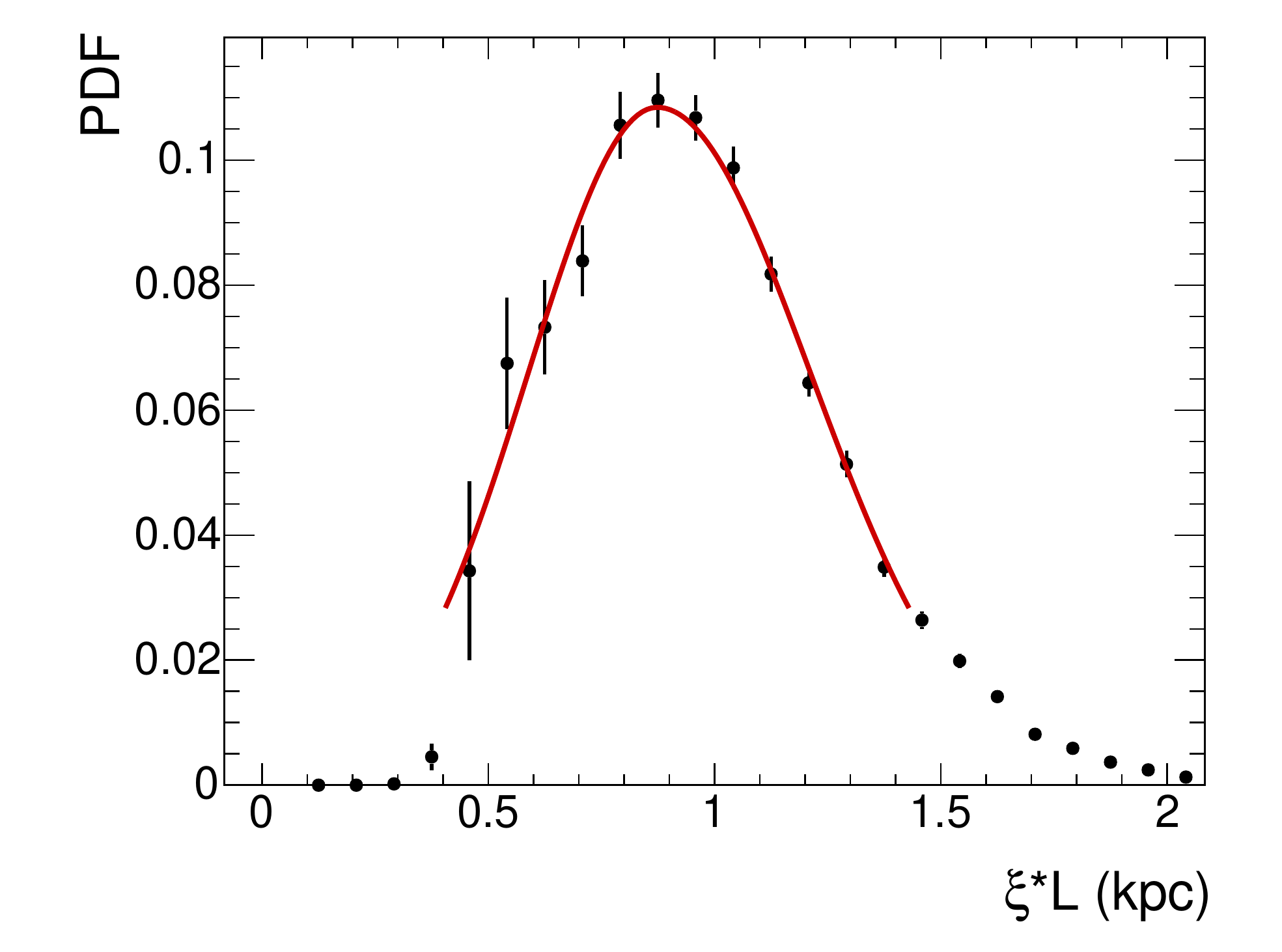}
  \caption{\captionsize%
    Top: two-dimensional contour plot for $\xi$$L$ and $D_{0}$.  
    The same color scale of Fig.\,\ref{Fig::ccContour} is used. 
    Bottom: posterior distribution for the parameter combination $\xi\,L$.
  }
  \label{Fig::xiL}
\end{figure}

A known problem in standard models of CR propagation is the parameter degeneracy between
the diffusion coefficient $D_{0}$ and the half-eight of the halo $L$.
In particular the parameter $L$ is of great importance for indirect searches of dark matter, because it
regulates the amount of dark-matter annihilation products that make up the CR flux.
This degeneracy can be in principle lifted with CR data on the radioactive isotope \BeTen.
For this reason we have included a large variety of \BeBe{} data in our dataset.
These data are shown in the bottom panel of Fig.\,\ref{Fig::NucleiSpectra} along with the 
THM calculations and corresponding uncertainty bands.
In contrast to other works that made use of similar sets of data \citep{PutzeMaurin:2010,Trotta:2011},
we found that under our model the parameter $L$ does not converge in probability.
As shown in Fig.\,\ref{Fig::ccContour}, the marginalized posterior distribution for the parameter $L$ is very broad.
This result can be easily understood in the context of the THM scenario,
because in this model the CR propagation region is split into two diffusive halos.
Hence, in comparison to standard OHM formulation, additional parameter degeneracies are expected
such as those involving $D_{0}$, $\chi$, $L$, and $\xi$.
Due to degeneracy, these parameters are correlated each other.
%
\begin{figure}[!t] 
  \includegraphics[width=0.46\textwidth]{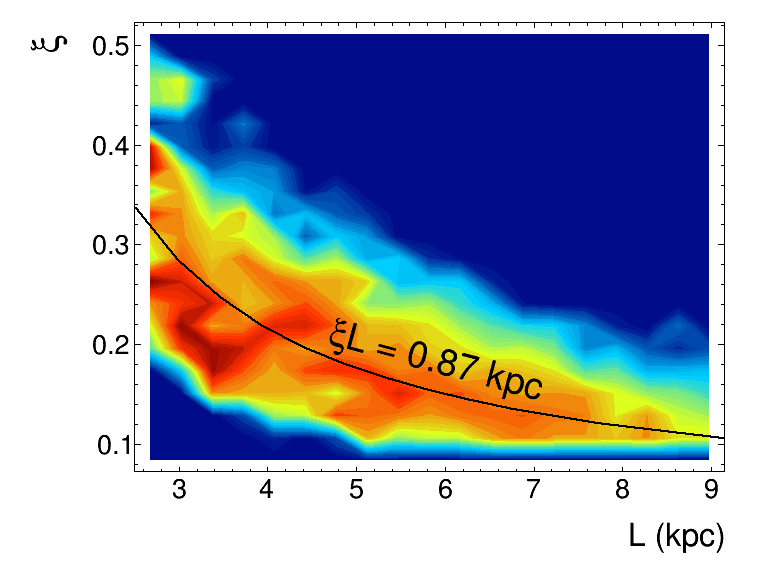}\caption{\captionsize
    Contour plot for the $\xi$ and $L$ plane. 
    The same color scale of Fig.\,\ref{Fig::ccContour} is used. 
    The black solid line corresponds to an inner-halo size $\xi \times L=0.87$\,kpc.}
  \label{Fig::xi_L}
\end{figure}
%
On the other hand, it is interesting to note that the half-height of the inner halo $\xi\,L$ is found to converge to about 0.9\,kpc. 
The correlation between  $\xi$ and $L$ is shown in the contour plot of Fig.\,\ref{Fig::xi_L},
where black solid curve represents the  product $\xi\,L=$\,0.87\,kpc.
In Fig.\,\ref{Fig::xiL} we directly plot the contour in the ($\xi\,L - D_0$) plane. It can be see that no significan correlation
is observed between $\xi\,L$ and $D_0$. Also, we have performed a fit on the $\xi\,L$ distribution in the interval
$\xi\,L\in \left[0.5, 1.5\right]$\,kpc using an asymmetric Gaussian function. 
The fit, shown in the plot of Fig.\,\ref{Fig::xiL}, describes the posterior distribution very well with $\chi^{2}/d.f. = 8.15/8$ 
and with a mean value of $\xi\,L = (0.87^{+0.34}_{-0.29})$\,kpc, which is also consistent with its best-fit value 0.85\,kpc.
%
These results can be explained as follows. The \BeTen{} isotope is unstable with lifetime $\tau_{0}=1.5$\,Myr.
Its diffusion scale distance is $\lambda \sim\sqrt{D \tau_{0}\gamma}$, where $\gamma$ is the Lorentz factor.
The existing observations are mostly gathered at sub-GeV/n kinetic energies, where one has $\lambda \lesssim \xi L$.
Thus the propagation of the \BeTen{} isotopes detected at Earth is essentially probing the inner halo diffusion properties. 
In this region the \BeTen{} equilibrium flux is approximately given by $J_{u}\sim\,Q_{s}\tau_{0}\gamma/\lambda$.
At low energy, the \BeBe{} ratio (or more in general, $u/s$ ratios between unstable and stable secondary nuclei)
is expected to behave as:
\begin{equation}\label{Eq::ApproximateTrendUnstable}
  J_{u}/J_{s} \sim \frac{\lambda(\R)}{\xi\,L} \propto \frac{\sqrt{D_{0}}}{\xi L} \,. 
\end{equation}
This relation, along with Eq.\,\ref{Eq::ApproximateTrendPrimary} and Eq.\,\ref{Eq::ApproximateTrendSecondary}, 
can explain why combined data on the \BC{} and \BeBe{} ratios may allow for the determination of $\xi\,L$.

It is also interesting to note that at increasing energies (from $\sim$\,1\,GeV/n to a few tenth of GeV/n), 
radioactive beryllium isotopes become more and more influenced by propagation properties in the outer halo.
Ideally, the collection of precise \BeBe{} data at $E \sim$\,5-50\,GeV/n would allow us to fully resolve the $L-\xi$
parameter degeneracy and eventually to resolve the single parameters $L$, $\xi$, $D_{0}$, or $\chi$.
Unfortunately this energy region is experimentally inaccessible by the current CR detection experiments. 
It is of particular interest, however, that the \AMS{} experiment is able to measure the \BeB{} elemental ratio from 0.5\,GeV/n to $\sim$\,1\,TeV/n.
As pointed out in several recent works \citep{PutzeMaurin:2010,TomassettiXS}, the use of the \BeB{} ratio in CRs
(in place of the \BeBe{} ratio) can enable us to recover the information contained in the \BeTen$\rightarrow$\BTen{} decay.   
While \AMS{} is expected to provide these data at the desired level of precision, a more serious limitation
is represented by cross-section uncertainties for the production of beryllium isotopes.
These uncertainties, in particular for the reaction $^{11}$\B+ISM\,$\rightarrow\,^{10}$\,\Be$+$\,X,
may affect our ability to resolutely break the parameter degeneracy, even with the availability of precise CR data \citep{TomassettiXS}.
Similarly, the solar modulation effect needs to be properly accounted. In this work, uncertainties in solar modulation
have been accounted conservatively as discussed in Sect.\,\ref{Sec::MCMC}.
These uncertainties are sub-dominant in comparison with the sizable error bars of the available CR data.
In the view of future precision data, a more reliable solar modulation modeling is probably necessary.

\subsection{Antiprotons}         
\label{Sec::ResultsAntiprotons}  

\begin{figure}[!t]
  \includegraphics[width=0.46\textwidth]{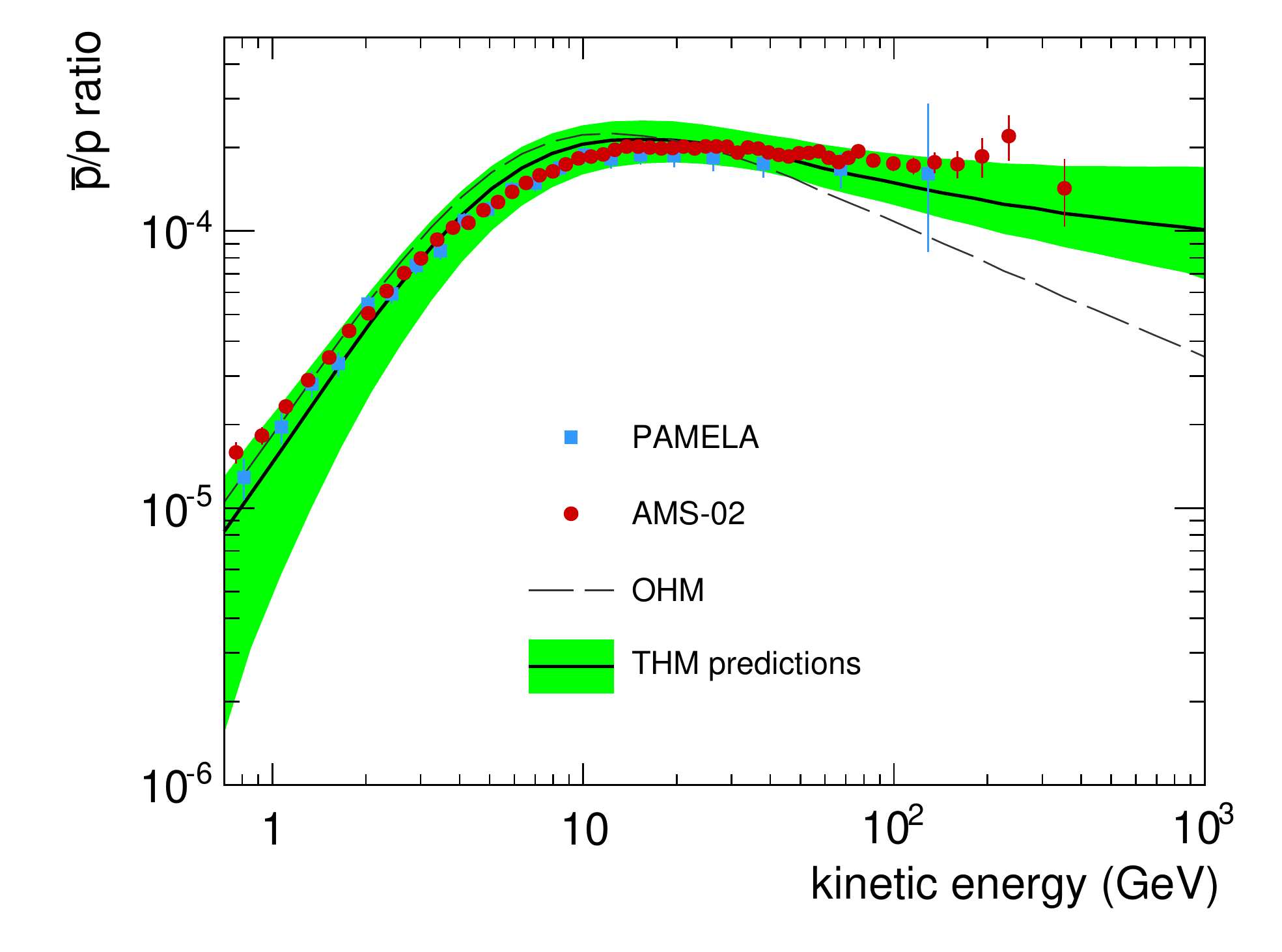}
  \includegraphics[width=0.46\textwidth]{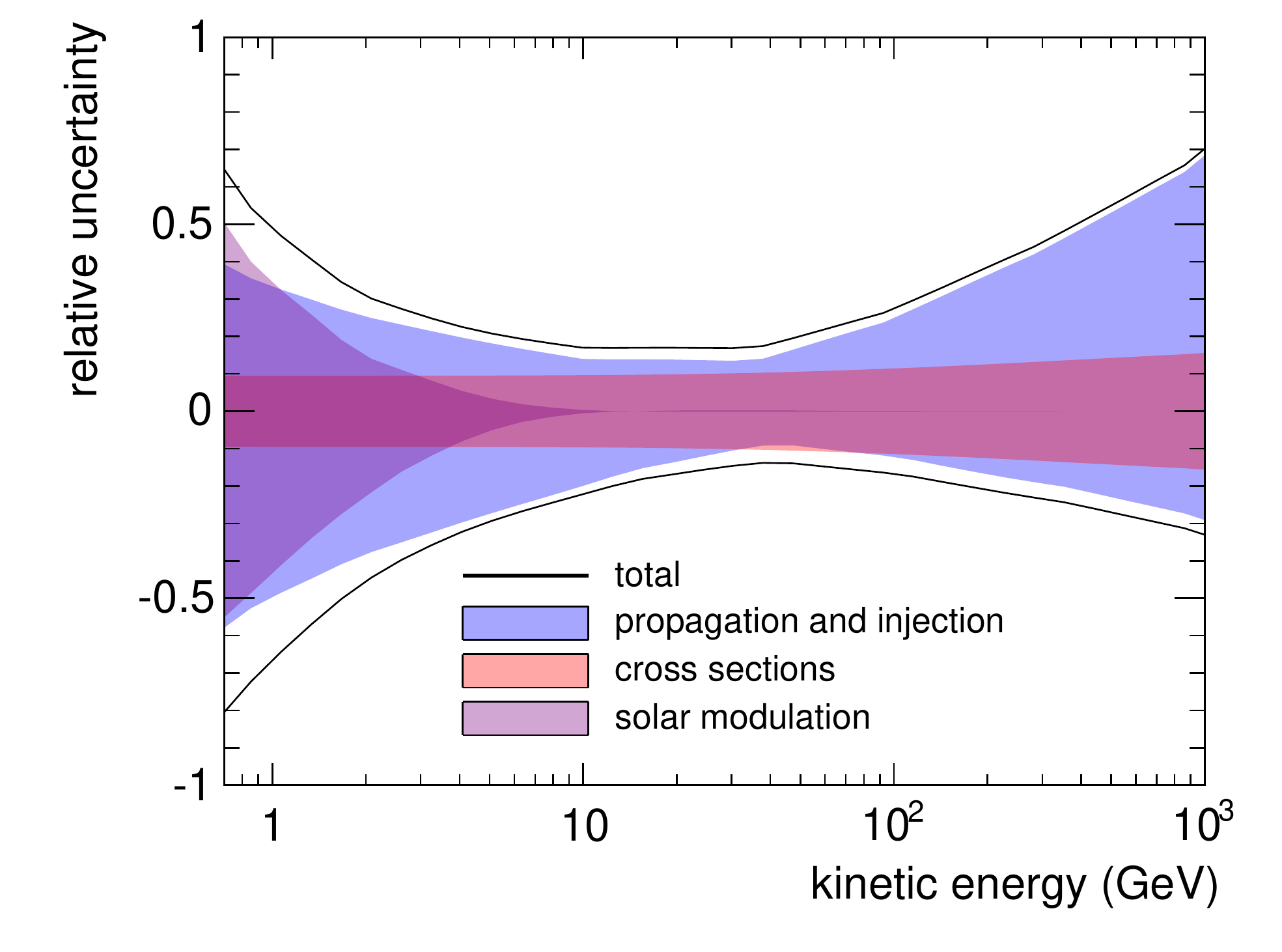}
  \caption{\captionsize%
    Top: \pbarp{} ratio as function of kinetic energy.
    Model calculations are shown in comparison with new data from \AMS{} \cite{Aguilar:2016pbar} and PAMELA \cite{PAMELA:Results}.
    The solar modulation level is set to $\phi=500$\,MV. The production cross-sections are evaluated using the MC generator \texttt{EPOS LHC} model \cite{Pierog:2013ria}.
    Bottom: estimated uncertainties for the \pbarp{} ratio arising from CR injection and propagation, production cross-sections, and solar modulations.
  }
  \label{Fig::ccAntiprotons}
\end{figure}

Antiprotons in CRs calculated from our models are of secondary origin and arise from collisions of
CR nuclei with the gas.
Using the MC generator \texttt{EPOS LHC}, we have evaluated the differential production cross-sections for all
reactions \p-\p, \p-\He, \He-\p, and \He-\He{} which generate antiprotons and antineutrons.
The latter decay rapidly in antiproton (with a rest lifetime $\tau_{0}\approx$\,15\,min),
therefore contributing appreciably to the observed flux.
Their subsequent transport of antiprotons in the Galaxy are similar to that of protons, apart from the
presence of \emph{non-annihilating} reactions of the type \pbar-\p$\rightarrow$\pbar$^{\prime}$-$X$
that produce an appreciable component of low-energy antiprotons.
In order to minimize systematic uncertainties in the model, in this work we make use of
the \pbarp{} ratio as key observable in place of the absolute antiproton flux.
The THM predictions for the \pbarp{} ratio are shown in the top panel of
Fig.\,\ref{Fig::ccAntiprotons} in comparisons with the new \AMS{} data \citep{Aguilar:2016pbar}.
Model calculations are referred to the best-fit parameter setting determined with the MCMC scan on nuclear data. 
The antiproton data of the figure are not directly included in the fit, as we are aimed at investigating the secondary nature of these particles.
According to our THM predictions, the ratio has a tendency to flatten at $E\sim$\,10\,GeV to $\sim$\, 100\,GeV.
This tendency is apparent in the comparison between THM and OHM model predictions, as the latter decreases steadily
at rigidity above $\sim$\,10\,GV. 
In comparison with secondary to primary nuclear ratios, the absence of a clear spectral change in the
\pbarp{} ratio is connected to particular shape of the antiproton source term.
In contrast to secondary matter nuclei, which carry about the same kinetic energy per nucleon of their progenitors, 
the antiproton production is characterized by broad energy distributions and large inelasticity factors. 
%
%
The green shaded band around the THM calculation includes various sources of uncertainty.
These uncertainties are reviewed in the bottom panel of the Fig.\,\ref{Fig::ccAntiprotons}, showing 
the contributions from injection and propagation, production cross-sections, and solar modulation. 
In principle, the CR propagation parameter uncertainties already included the solar modulation 
uncertainties, because they have been accounted in the MCMC procedure. 
However, charge-sign and mass dependent solar modulation effects are in general expected due 
to particle drift or adiabatic losses of CRs in the heliosphere, that are unaccounted by the force-field model.
Hence the use of CR proton data does not provide safe constraints on the solar modulation of antiprotons.
Following \citet{Giesen:2015ufa}, we have varied the solar modulation potential from 200 MeV to 700 MeV to estimate this error.
This estimate encompasses the level modulation asymmetry between protons and antiprotons, that we have tested using the model of \citet{CholisHooper:2015}. 
The solar modulation error is dominant at 1 GeV/n of energy and becomes negligible at 15 GeV/n in comparison with the uncertainties of the experimental data.
A large uncertainty factor comes from antiproton production cross-sections. The figure shows that the cross-section contribution is 10\% at 1 
GeV/c and increases slowly with energy to become 18\% at about 1 TeV/c. The calculations of these errors can be found in Appendix\,\ref{Sec::ApppendixB}.
In the high-energy region, errors are dominated by uncertainties in CR injection and propagation parameters. 
In contrast to other works \cite{Giesen:2015ufa}, our fitting procedure lead to a unique astrophysical uncertainty factor which include the
errors from propagation effects and those induced by primary nuclei. However no appreciable correlation is found the two contributions. 
At kinetic energy above $\sim$\,100\,GeV, this uncertainty is at the level of $\sim30\%$ and it is limited by the experimental errors
of the high-energy \BC{} ratio. Parameters describing CR injection spectra of protons and \He{} are better constrained with the existing
data, although their contribution to the total \pbarp{} uncertainty band becomes non-negligible at high energies.
In summary, under our scenario of spatial-dependent CR propagation, the predictions for the \pbarp{} ratio appear to be fairly consistent with the \AMS{} data,
within the present level of uncertainty, showing no striking evidence for an antiproton excess. 
We note, nevertheless, that the dominant contribution to the uncertainties is related to CR propagation. Hence the situation
will become more transparent with the availability of precise \BC{} data at TeV/n energies \citep{Maestro:2015ICRC}.

\subsection{Positrons}        
\label{Sec::ResultsPositrons} 

\begin{figure}[!t]
  \includegraphics[width=0.46\textwidth]{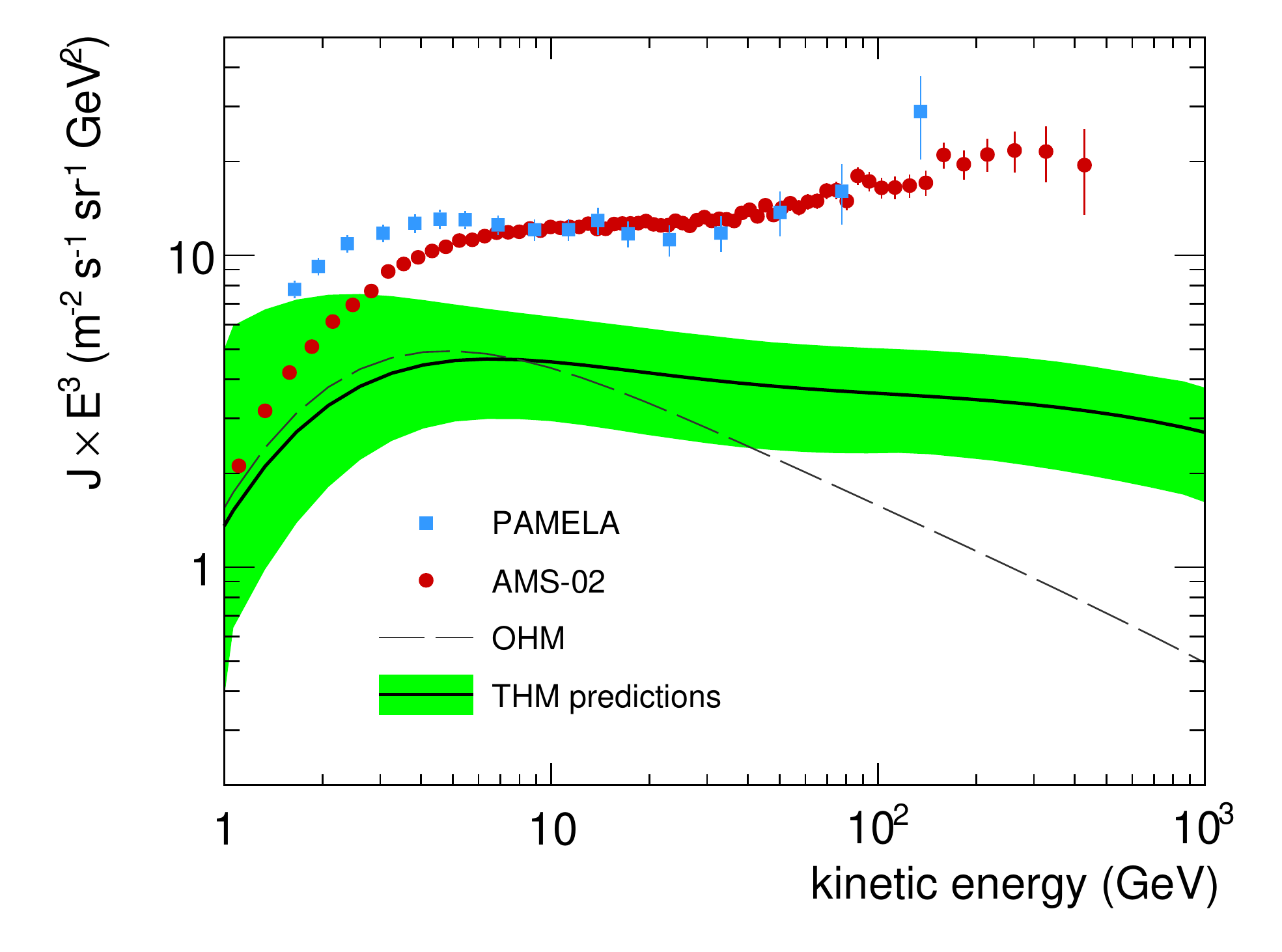}
  \includegraphics[width=0.46\textwidth]{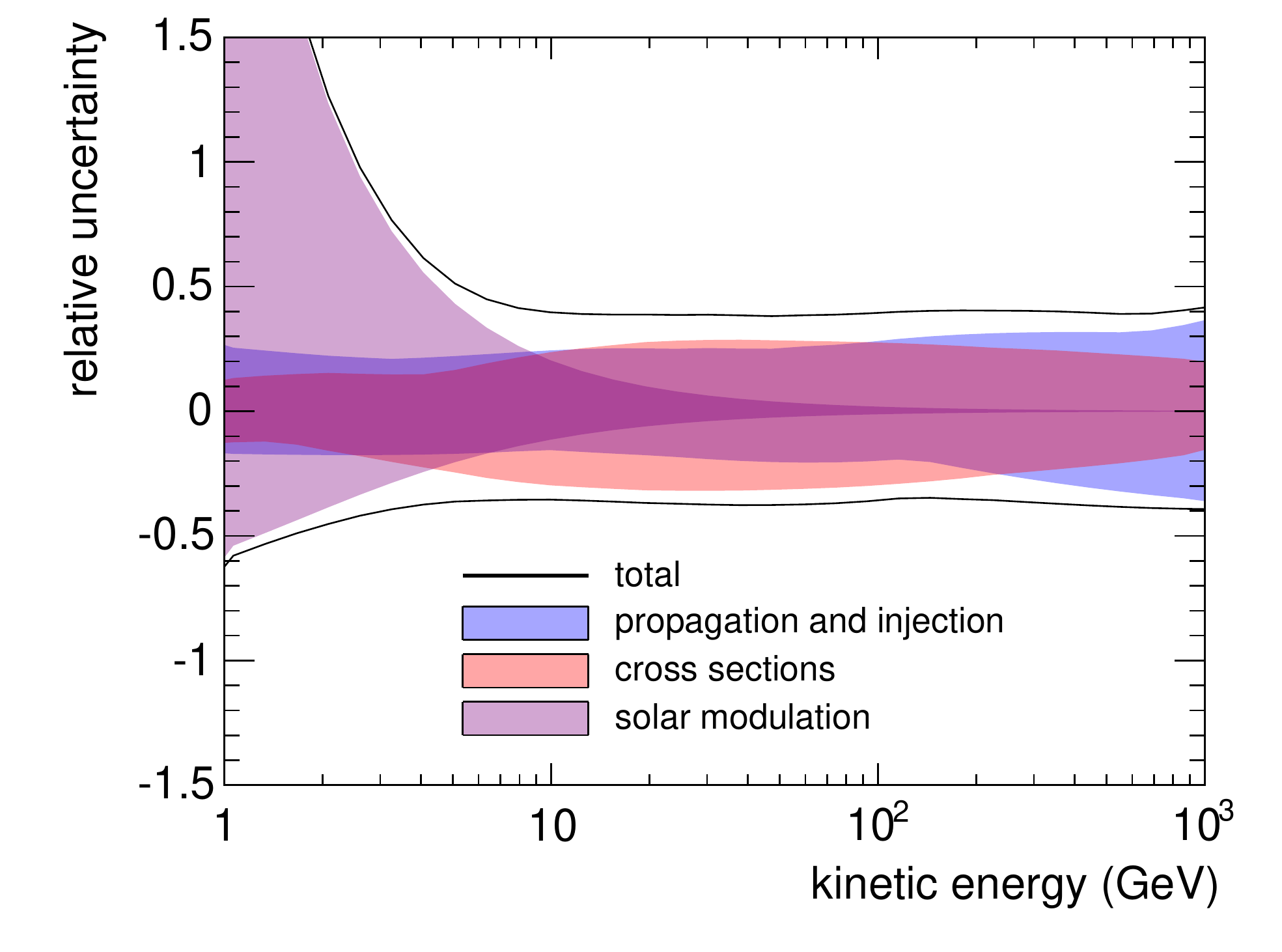}
  \caption{\captionsize%
    Top: secondary positron flux in CRs as function of kinetic energy.
    Model calculations are shown in comparison with the data from  \AMS\ \cite{AMS02:Leptons} and PAMELA \cite{PAMELA:Results}.
    The solar modulation level is set to $\phi=500$\,MV.  
    Bottom: estimated uncertainties for the positron flux arising from CR injection and propagation, production cross-sections, and solar modulations.}
  \label{Fig::ccPositrons}
\end{figure}

Similarly to antiprotons, secondary positrons are generated by collisions of CR hadrons with the ISM. 
Thus we consider the absolute flux of CR positrons rather than positron fraction \epfrac,
because it permits to avoid further assumptions on the injection spectrum of primary electrons.
The predicted flux of secondary positrons is shown in the top panel of Fig.\,\ref{Fig::ccPositrons}.
The black solid line represents the THM model calculations under the best-fit parameter set, while the
shaded band is the corresponding total uncertainty.
%
%
The positron flux predicted by our model is significantly harder than that arising from the OHM setting.
The main reason for this difference is that CR positrons detected at Earth have spent a large fraction 
of their propagation time in the region close to the Galactic disk.
Given the shallow diffusion of CRs in the inner halo, the flux steepening effect induced by
diffusive propagation is expected to be milder for the THM model, in comparison with standard OHM calculations.
In addiction to diffusive propagation, however, energy losses arising from synchrotron radiation and
inverse Compton processes have an important impact in reshaping the spectrum of charged leptons.
For these effects, the energy loss rate is of the type $b(E)=b_{0}E^{2}$, with $b_{0}\cong\,1.4\times10^{-16}$\,{GeV}$^{-1}$\,{s}$^{-1}$ \cite{Leptons:Pulsar}.
The time-scale of these processes is $\tau=\left(b_{0}E\right)^{-1}$, so that the typical diffusion 
scale distance is of the order of $\lambda \sim \sqrt{\tau\,D}$.
More precisely, for the propagation of CR electrons and positrons from the Galactic disk, one can write
\begin{equation}
  \lambda(E,E_0)= 2\left\{\frac{D_{0}E^{\delta}}{b_{0}E(1-\delta)} \left[ 1-\left(\frac{E_{0}}{E}\right)^{\delta-1}\right]\right\}^{\frac{1}{2}} \,,
\end{equation}
where $E_{0}$ is their initial energy.
For detected positron energy $E$ in the $\mathcal{O}$(100\,GeV) energy scale and $E_{0}\gtrsim\,E$,
it can be seen that the diffusion distance $\lambda$ is always $\lesssim\,1$\,kpc for our best-fit propagation parameters.
Hence the propagation histories of high-energy positrons detected at Earth take place essentially in the inner halo.
In this region, the CR positron fluxes are of the type $J_{+}\sim (\tau/D)^{1/2}Q^{\rm sec}$ so that,
for proton-induced source spectra $Q^{\rm sec}\sim\,E^{-\gamma_{p}}$, one has $J_{+}\propto\, E^{-\gamma_{p}-\frac{1}{2}(\delta+1)}$.
Note also that, for $E_{0}\gg\,E$ and in particular for $E\lesssim\,10\,$GeV, the quantity $\lambda(E,E_{0})$ can reach larger values.
Thus, in the general cases, CR leptons may experience propagation in both halos and their resulting flux at Earth is a convolution
over their propagation histories. 

In the bottom panel of Fig.\,\ref{Fig::ccPositrons} we provide a breakdown of the main
sources of uncertainties associated with the positron flux calculations.
The errors on the production cross-sections are estimated as in \citet{Delahaye:2008ua},
\ie, by evaluating the effects of different cross-section parameterizations as a function of energy.
The considered parameterizations are those proposed by
\citet{Kamae:2006}, \citet{TanNg:1983}, and \citet{Badhwar:1977}.
The positron source term is found to vary between 5$\%$ ~ 30$\%$ with energy, depending on the adopted parameterization. 
The uncertainties of solar modulation are estimated by varying the modulation potential $\phi$ similarly to the antiproton
case of Sect.\,\ref{Sec::ResultsAntiprotons}. In comparison to other source of uncertainties, solar modulation uncertainties
are important below 10 GeV. In comparison with the experimental errors of \AMS{} measurements, they become negligible above a few tens GeV. 
Uncertainty from CR propagation and injections are those estimated by the MCMC parameter scan procedure.
It is worth pointing out that the positron flux is still softer than $E^{-3}$ while the data measured by \AMS{} is harder.
To account for the missing flux, it is necessary to add some extra contribution of high-energy positrons.
Primary positron sources may include nearby pulsars, old SNRs or dark matter particle annihilation.
They are preferentially located within relatively short distances.

\subsection{Anisotropy}        
\label{Sec::ResultsAnisotropy} 

With the THM parameter setting of the best-fit configuration, we have calculated the flux
anisotropy amplitude at the location of the Sun due to global leakage of CRs from the Galaxy. 
%
\begin{figure}[!t]
  \includegraphics[width=0.46\textwidth]{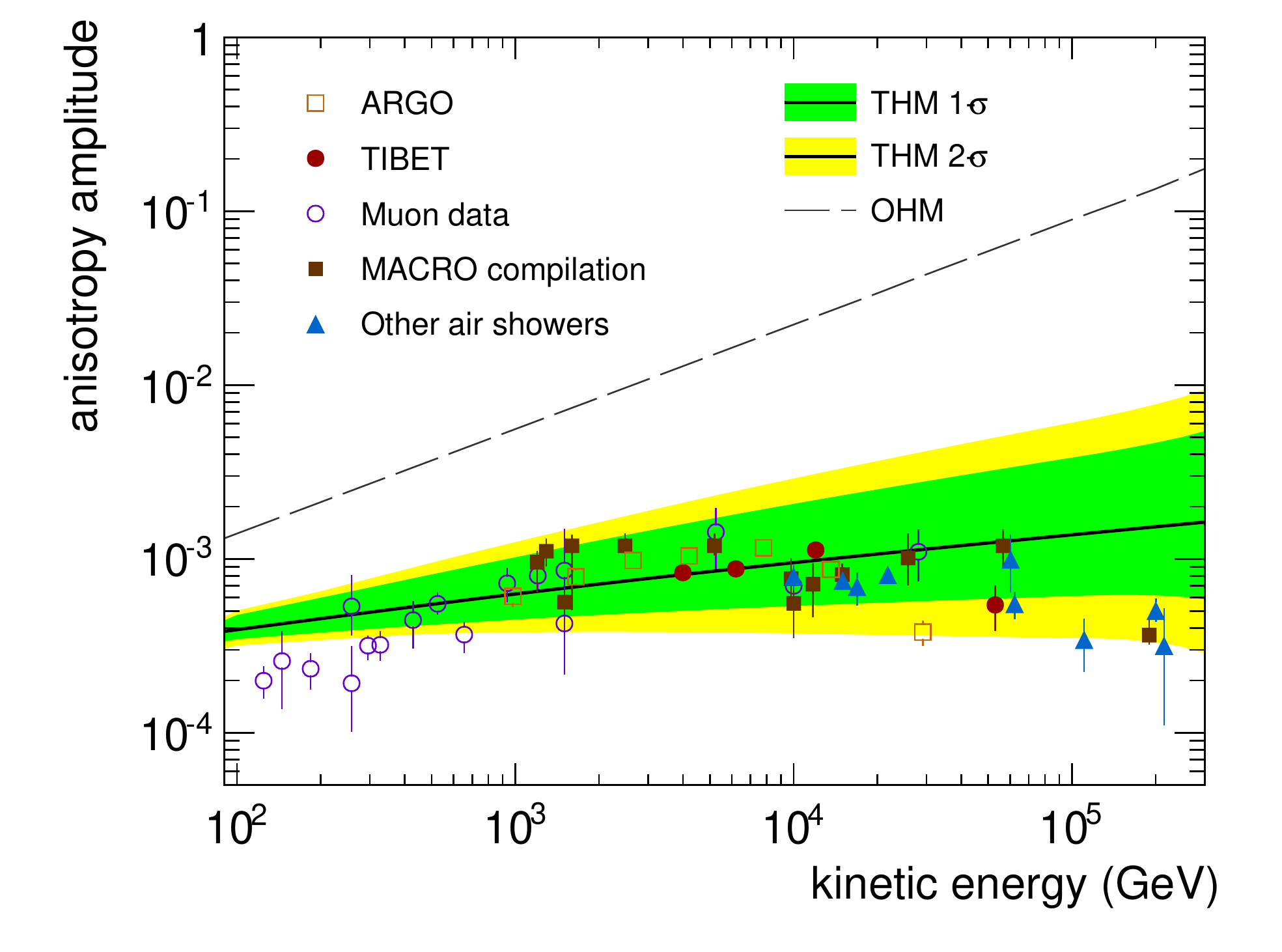}
  \caption{\captionsize%
    Anisotropy amplitude from best-fit THM calculations in comparisons with the data at $E\approx$\,100\,GeV - 300\,TeV. OHM calculations are shown for reference.
  }
  \label{Fig::ccAnisotropy}
\end{figure}
%
In the diffusion approximation, the anisotropy is dominated by the radial streaming of the CR flux. 
Its amplitude $\hat{A}$ is computed as:
\begin{equation}\label{Eq::Anisotropy}
  \hat{A} = \frac{3\, D(\R)}{c\, \mathcal{\psi}} \left| {\nabla\mathcal{\psi}}\right| \,.
\end{equation}
The calculations are shown in Fig.\,\ref{Fig::ccAnisotropy} in comparison with a compilation of data at energy $E\sim$\,100\,GeV - 100\,TeV. 
The high degree of isotropy observed in CRs cannot be accounted by standard models of CR propagation, which suffer from a too fast increase
of the anisotropy amplitude at high-energy. In contrast, under our model, the predicted anisotropy is found smaller and less energy dependent
than that generally predicted from standard diffusion model \citep{CR:Review}.
From Eq.\,\ref{Eq::Anisotropy}, the quantity $\hat{A}$ is highly sensitive to the rigidity dependence of the \emph{local} diffusion
coefficient, which in our model is found to increase as weakly as $\sim\R^{0.15}$. Furthermore, the slow diffusion of CRs in the
inner halo produces a small radial gradient of the total flux, which contribute in lowering the overall anisotropy amplitude. 
In Fig.\,\ref{Fig::ccAnisotropy}, the uncertainty band arise from the knowledge on the total CR flux and
on the diffusion coefficient. While the former is well constrained by proton and \He{} flux data, the latter 
suffers from experimental uncertainties in the multi-TeV \BC{} ratio.
Hence the error band of Fig.\,\ref{Fig::ccAnisotropy} reflects, essentially, the uncertainties on the \BC{} ratio. 

Although the model agreement with the data is very good, we stress that we do not include anisotropy data in the likelihood,
because there are other possibilities to explain the observations that are unaccounted by our model \citep{Kachelriess2015SNR}. 
For example, the level of anisotropy can be significantly affected by the presence of nearby and localized sources of CRs, which is ignored in our calculations. 
Calculations would depend on the assumed distances to the closest sources and their ages, but the precise distribution of sources 
is essentially unknown.
Another possibility is having a different (smaller) diffusion coefficient $D$ in the very local environment around the
solar system, \ie, inside the \emph{local bubble}, that would effectively isotropize the locally observed flux.

\section{Discussion}    
\label{Sec::Discussion} 

In this section we make some considerations about our findings and discuss some limitations of our study.
We find that, overall, a THM scenario of diffusive propagation describes very well the observed properties of the CR spectrum.
The idea of having two diffusive halos represent the simplest and physically consistent generalization of the standard CR propagation models, 
as the latter assume that a unique diffusion regime is at work everywhere.
The origin of the two zones relies in the link between CR diffusion and interstellar turbulence, but this link 
is not self-consistently addressed under our phenomenological implementation.
Hence we have no \emph{a priori} prescription for the parameters describing spatial extents or diffusion properties of these regions,
and we used the data for constraining their values. In this respect, the MCMC method of parameter sampling is found to be well suited for the purpose. 

In all plots, OHM calculations are given for comparison purposes. This model is intentionally tuned to
data below 200\,GeV/nucleon energies, thus it underpredicts the data at higher energies.
To solve this problem, several recent models based on homogeneous diffusion make use of broken
injection \citep{Korsmeier2016,Johannesson:2016,Giesen:2015ufa,Pbar:Unc}. With a break in the
injection spectrum at $\R\sim\,300$\,GV of rigidity, the data on primary CR fluxes can be described well.
This assumption implicitly ascribes the observed spectral hardening as an intrinsic property of CR sources, 
which might find explanations in the time-dependent nature of DSA \citep{OhiraMalkov} or in
the feedback of accelerated particles on the SNR shocks \citep{Ptuskin2013}. 
Under these models, secondary-to-primary ratios do not show any tendency to flatten and
the \BC{} ratio decreases rather sharply at high energy (not differently to the OHM line of Fig.\,\ref{Fig::NucleiSpectra}).
The incorporation of diffusive reacceleration in CR transport may allow for a better description of the current 
\BC{} data, including the observed peak at $\sim$\,1\,GeV/n energy.
However, it requires questionably large values for the Alfv\'enic speed (of the order of $v_{A}\approx$\,30-40\,km/s) and
the introduction of additional injection breaks at $\R\sim$\,10\,GV to avoid the production of unphysical bumps in primary spectra.
A striking difference between the two models, OHM and THM, is their prediction for secondary antimatter production.
Given the role of antiprotons in dark matter searches, the discrimination among the two scenarios is of crucial importance.
Accurate measurements of the \BC{} or \LiC{} ratios at the TeV energy scale will be hopefully
able to achieve such a discrimination \citep{Maestro:2015ICRC,Serpico:2015ICRC}. 

There are some important simplifications in our calculations that we briefly discuss.
First, our work is based on the usual picture that the CR flux arises from a large population of contribution SNRs.
These SNRs are modeled as steady-state and continuously distributed on the Galactic plane.
However, the stochastic nature of SNR events may induce deviations in our predictions. In particular the 
anisotropy amplitude is critically sensitive to nearby-source contributions to the flux. 
Also, the presence of CR accelerators in our local environment may also appear in primary or secondary 
spectra \citep{Kachelriess2015SNR,TomassettiDonato,Tomassetti2015TwoSnr}.
A second simplification concerns the properties of the very local interstellar medium. 
The solar system is situated inside a $\sim$\,200\,pc-sized underdense environment, the local bubble, which
may influence the production of radioactive isotopes such as \BeTen{} or $^{27}$\Al{} \citep{PutzeMaurin:2010}. 
In this work, data on the \BeBe{} ratio play a key role in determining the properties of the inner halo.
These results rely on the simplified hypothesis that the interstellar gas is smoothly distributed on the Galactic plane.
Also, a precise interpretation of the \BeBe{} ratio may require a better modeling of the solar modulation effect.
Given the large error bars of the existing data, we found that the uncertainties arising from solar modulation
are not critical, hence we adopted a simple conservative approach.
With the availability of precision data on \BeBe{} or \BeB{} ratios, an improved modeling will be required on this side. 

We also note that the fitting procedure of Sect.\,\ref{Sec::MCMC} requires that all errors that make up the likelihood are uncorrelated with each other. 
This assumption breaks down for high-statistics measurements dominated by systematic errors. For these experiments, however,
bin-to-bin correlations matrices are not available. Hence we may have slightly over-estimated the THM uncertainties on primary CRs. 
On the other hand, in the model predictions of antiparticle spectra, the systematic uncertainties from solar modulation
and production cross-sections have been treated separately and eventually summed in quadrature. Since no Gaussian behavior
is expected from these uncertainties, and their PDFs are unknown, we did not define 1-$\sigma$ and 2-$\sigma$
confidence levels for these errors.

\section{Conclusions}     
\label{Sec::Conclusions}  
        
Understanding acceleration and propagation processes of CRs in the Galaxy is a central question in astrophysics.
In this work, we have performed a large scan on the CR injection and transport parameter space
using a spatial-dependent model of CR diffusive propagation. 
More specifically, we set up a numerical implementation of a two-halo model where
CR propagation takes place
in two regions characterized by different energy dependences of the diffusion coefficient.
As shown, such a model is able to account for several observed properties of Galactic CRs
such as the energy spectra of primary nuclei, the shape of secondary-to-primary ratios,
the relative abundance of radioactive isotopes, and the high-degree of flux isotropy.
In particular, a distinctive feature of this model is a high-energy hardening of secondary to primary ratios.
Using a large variety of CR data, and a selected set of key astrophysical parameters,
we have performed a global Bayesian analysis based on a MCMC sampling technique.
With this method we obtained the marginalized probability density functions for several free parameters 
while properly accounting for their correlations.

By testing our model of CR propagation using a large set of experimental data, 
we have inferred that the inner halo surrounds the Galactic plane for a vertical extent of about  $\xi\,L\,\approx$\,900\,pc.
The estimated size of this region supports the conception that SN explosions are the source of magnetic turbulence
in the Galactic disk while in the far outer halo, where the SNR activity is reduced, the turbulence is driven by CRs themselves. 
According to our estimates, the inner halo is characterized by a small dependence for the diffusion coefficient with
rigidity, $D\propto \R^{0.15}$. From this dependence, the high-energy flux of Galactic CRs near the solar system results
in a high degree of flux isotropy, with amplitude $\hat{A}\approx\,10^{-3}$ in the multi-TeV energy region,
in nice accordance with observations.

A prime goal of our work is to perform an evaluation of the astrophysical antimatter background
and its uncertainties in the highest measured energy region.
We have assessed the uncertainties associated with the main ingredients of the calculations
of antiproton and positron production from collisions of CRs with the gas.
Astrophysical uncertainties from injection and propagation have been obtained from our MCMC global scan 
in terms of 
$1-\sigma$ and $2-\sigma$ envelopes on the MCMC output for antiproton and positron fluxes.
Other relevant systematic errors are those associated with production cross-sections.
To evaluate antiproton production cross-sections and corresponding uncertainties,
we made use of hadronic MC generators in combination with recent \p-\p{} and \pC{} data from high-energy collision experiments.                   
We have eventually presented the model predictions for the positron flux and the antiproton/proton
ratio between $\sim$\,0.5 and $\sim$\,1000\,GeV of kinetic energy. 
Our secondary production calculations for the CR positron flux, in spite of large uncertainties in the modeling,
cannot account for the data without the introduction of some extra component of primary positrons.
In contrast, the new $\pbarp$ ratio reported by \AMS{} is fairly well described by our calculations within the estimated errors. 
We obtained an improved level of level of agreement in comparison with other recent works
in which the preliminary \AMS{} data were found to lie at the upper edge of the uncertainty band. 
It has been noted, in fact, that a mere astrophysical interpretation of the \AMS{} antiproton data demands a weak 
energy dependence for the CR diffusion at high energy. But this tension cannot be satisfactorily resolved with 
standard diffusion models given the observed decrease of the \BC{} ratio. In comparison with conventional models of CR propagation,
secondary antiparticle fluxes from our calculations are considerably enhanced at the highest detectable energies.
For both species, and in particuar for antiprotons, uncertainties arising from CR propagation are still the dominant contribution at high-energy.
In the near future, with the availability of more precise \BC{} data from \AMS{}, CALET or DAMPE,
all uncertainties related to the modeling of CR propagation are expected to be dramatically reduced,
so that nuclear uncertainties will represent a major limitation for the interpretation of CR data.
Thus we consider critical to improve the cross-section measurements
and we hope that ongoing particle physics experiments at LHC will provide valuable data on antiproton production.

\section*{Acknowledgments}  

We thank our colleagues of the AMS Collaboration for valuable discussions,
and the \Dragon{} team for sharing their code with the community.
We are indebted to Tanguy Pierog, Colin Baus, and Ralf Ulrich for the \CRMC{} interface to access the MC generators.
We are grateful for important discussions with Anatoly Erlykin, Sujie Lin, Sergey Ostapchenko and Arnold Wolfendale.
JF acknowledges the support from China Scholarship Council and the Taiwanese Ministry of Science and
Technology (MOST) under Grant No. 104-2112-M-001-024 and Grant No. 105-2112-M-001-003.
AO acknowledges CIEMAT, CDTI and SEIDI MINECO under Grants ESP2015-71662-C2-(1-P) and MDM-2015-0509.
   NT acknowledges support from MAtISSE - \emph{Multichannel Investigation of Solar Modulation Effects in Galactic Cosmic Rays}.
This project has received funding from the European Union's Horizon 2020 research and innovation programme under the Marie Sklodowska-Curie grant agreement No 707543.


\appendix

\section{models of \pbar{} production cross-sections}  
\label{Sec::ApppendixB}                                

In the past 20 years, antiproton production cross-sections parameterized by \citet{TanNg:1983} have been widely used in CR propagation calculations. 
Due to the lack of high-energy measurements, this semi-empirical parameterization has been tuned to the experimental data avaliable at the epoch
and then extrapolated to the relevant energies. Recently, high-energy collision experiments have triggered some 
efforts in updating the antiproton cross-section models \cite{diMauro:2014zea,Kachelriess:2015wpa,Kappl:2014}. 

In this Appendix, we make use of recent data on antiproton production from \pp{} and \pC{} 
collisions reported by
NA49\,\cite{NA49}, BRAHMS\,\cite{Arsene:2007jd}, and ALICE\,\cite{Aamodt:2011zj}, 
to constrain the cross-section calculations of MC generators such as 
\texttt{EPOS LHC}, \texttt{EPOS 1.99}~\cite{Pierog:2013ria}, \texttt{SIBYLL}~\cite{Engel:1999db}, and \texttt{QGSJET-II-04}~\cite{Ostapchenko:2010vb}.
These MC generators are widely used in the simulation of extensive CR air showers and have
been recently tuned to reproduce minimum bias LHC Run-1 data \cite{Pierog:2013ria,Ostapchenko:2010vb}.

\begin{figure}[!t]
\includegraphics[width=0.46\textwidth]{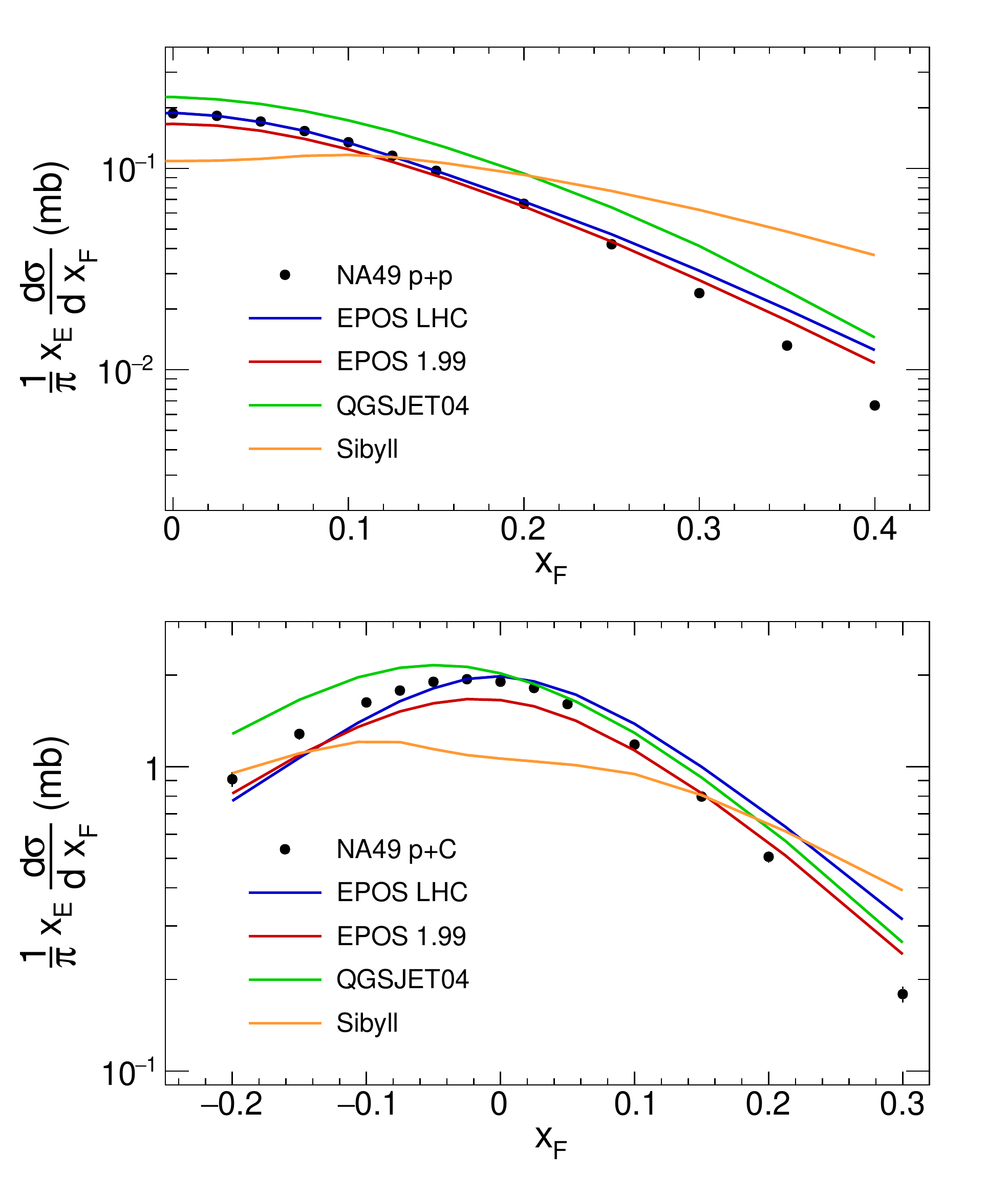}
\caption{
Feynman $x$-spectra of antiprotons, in \pp{} (top) and \pC{} (bottom) collisions measured by the NA49 experiment\,\cite{NA49} (dots).
Corresponding calculation from MC generators are superimposed (lines).
}
\label{Fig::na49_integrated}
\end{figure}

\begin{figure}[!t]
\includegraphics[width=0.46\textwidth]{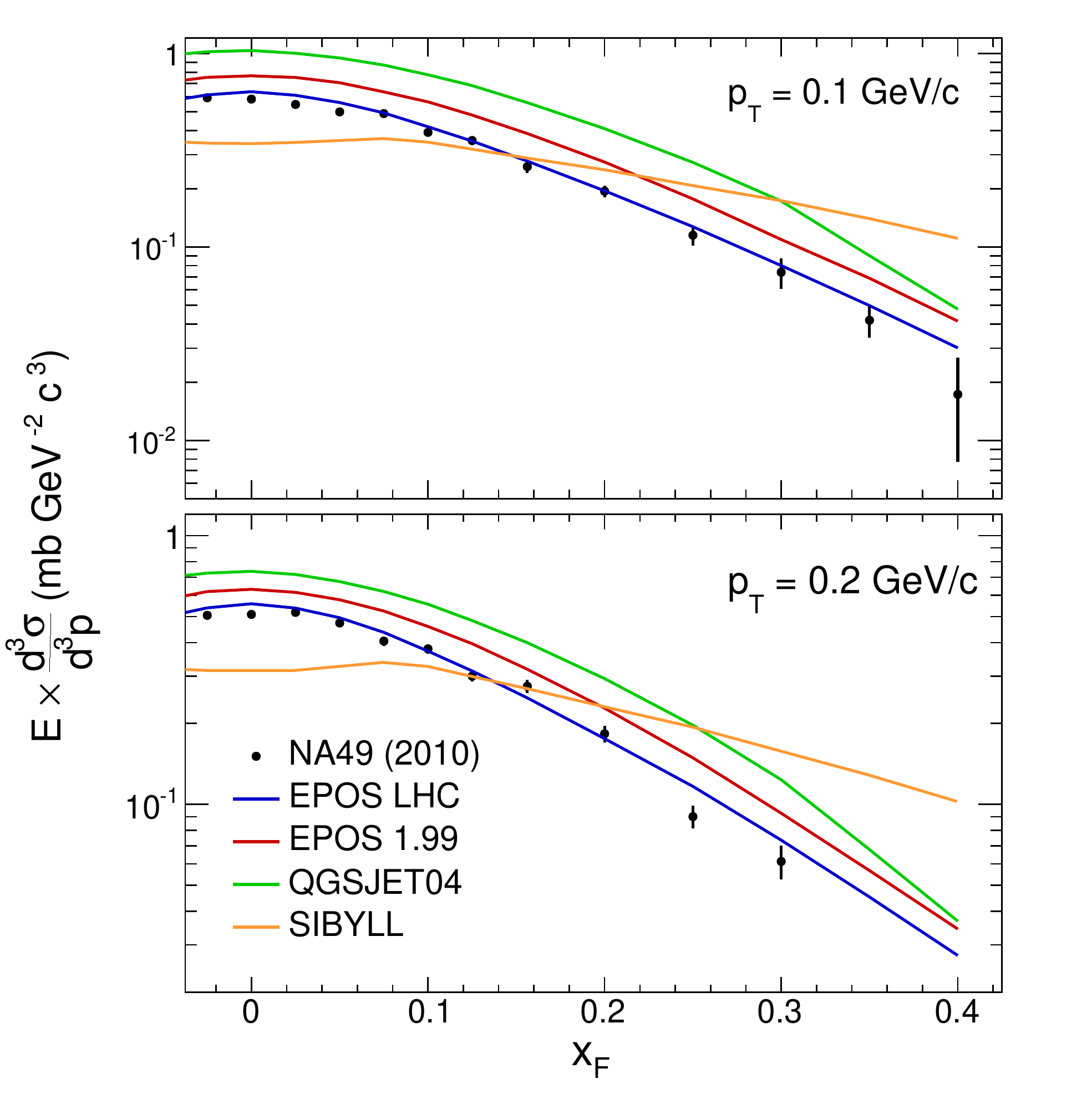}
\caption{
The \pp{}$\rightarrow$\,\pbar{} production invariant cross-section measured by the NA49 experiment~\cite{NA49} (dots) 
for two transverse momenta  $p_{T} = 0.1, 0.2$ GeV/c (top, bottom), as function of the Feynman-$x$ variable $x_{F}$. 
Corresponding calculation from MC generators are superimposed (lines).
}
\label{Fig::na49_pp}
\end{figure}

\begin{figure}[!t]
\includegraphics[width=0.46\textwidth]{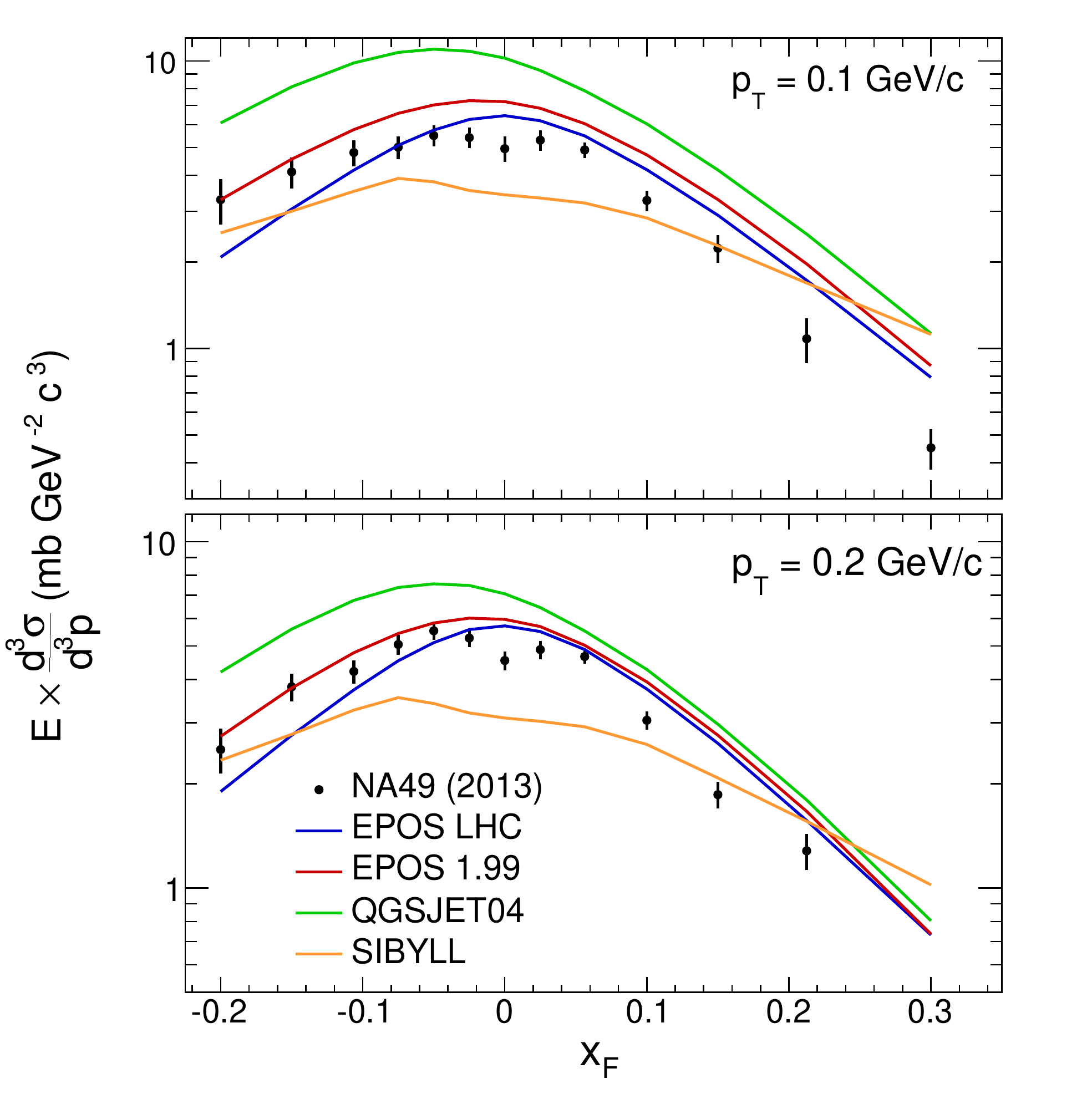}
\caption{
The \pC{}$\rightarrow$\,\pbar{} production invariant cross-section measured by the NA49 experiment~\cite{NA49} (dots) 
for two transverse momenta  $p_{T} = 0.1, 0.2$ GeV/c (top, bottom), as function of the Feynman-$x$ variable $x_{F}$. 
Corresponding calculation from MC generators are superimposed (lines).
}
\label{Fig::na49_pC}
\end{figure}

Figures\,\ref{Fig::na49_integrated},~\ref{Fig::na49_pp} and~\ref{Fig::na49_pC} show the \pbar{} production cross-sections measured by the NA49 experiment at CERN.
In this experiment, antiprotons are generated by
a 158 GeV/c momentum proton beam extracted at the Super Proton Synchrotron interacting on H or C steady targets.
Figure\,\ref{Fig::na49_integrated} shows the Feynman $x$-spectra of antiprotons $x_{E}/\pi\,(\textrm{d}\sigma \, /\textrm{d}x_{\textrm{F}})$ as function 
of the Feynman scaling variable $x_{F}$, where $x_{E} =  2E^{*}/\sqrt{s}$ and $x_{\textrm{F}} =  2p^{*}_{L}/\sqrt{s}$, 
being $E^{*}$ and $p^{*}_{L}$ the \pbar{} energy and momentum longitudinal component in the center of mass frame. 
Corresponding calculation from MC generators are superimposed. It can be seen that \texttt{EPOS LHC} performs slightly better than other models. 
Figures\,\ref{Fig::na49_pp} and\,\ref{Fig::na49_pC} present the NA49 $\bar{p}$ production invariant cross-section for two different transverse 
momenta, $p_{T}=$\,0.1 and 0.2 GeV/c.
The agreement with the \texttt{EPOS LHC} generator if fairly good, also for the dependence with transverse momentum.

\begin{figure}[!t]
\includegraphics[width=0.46\textwidth]{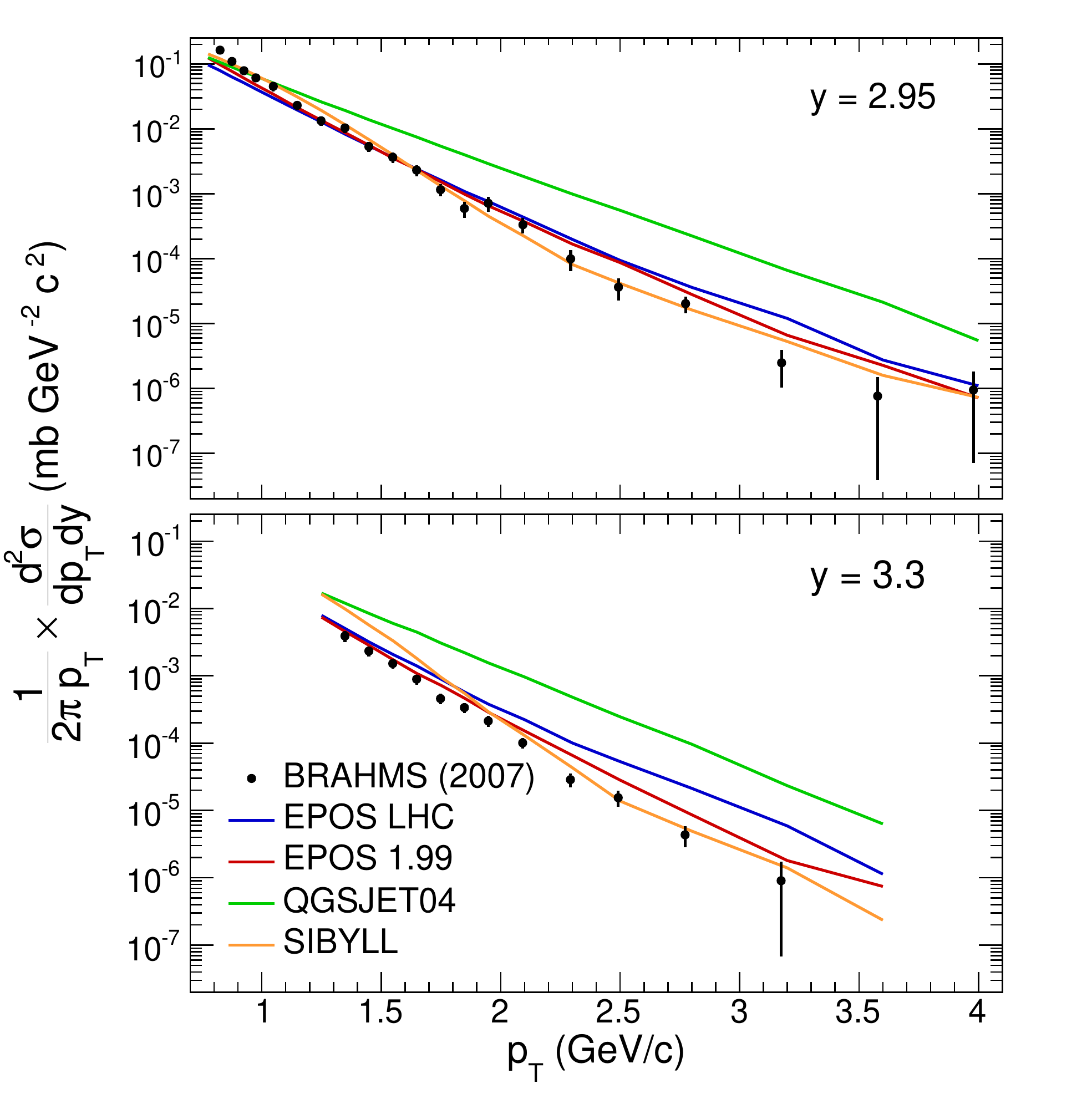}
\caption{
The \pp{}$\rightarrow$\,\pbar{} production invariant cross-section measured by the BRAHMS collaboration~\cite{Arsene:2007jd} (dots) 
for two rapidities $y = 2.95, 3.3$ (top, bottom) as function of $p_{T}$. 
Corresponding calculation from MC generators are superimposed (lines). 
}
\label{Fig::brahms_pp}
\end{figure}

Figure\,\ref{Fig::brahms_pp} shows a similar measurement reported by the BRAMHS collaboration. 
The measurement is operated the RHIC collider, Brookhaven National Laboratory, by \pp{} interactions at center of mass energy $\sqrt{s} = 200$ GeV.
Data are displayed for two different values for the rapidity $y$, as function of $p_{T}$ and compared to simulations. 
Rapidity is defined as:
\begin{equation}\label{Eq::Rapidity}
y = \frac{1}{2} \ln \left( \frac{E^{\star}+p^{\star}_{L}}{E^{\star}-p^{\star}_{L}} \right) \,.
\end{equation}
The most accurate description of the data comes from \texttt{EPOS 1.99}.
These data are collected at relatively high $p_{T}$, in a region that has little impact for the 
$\bar{p}$ secondary production in cosmic rays. 
However, BRAHMS data contribute to the understanding of the overall MC model validity and to 
the determination of the uncertainties to be associated with the MC model in an energy range where 
no other data are available.

\begin{figure}[!t]
\includegraphics[width=0.46\textwidth]{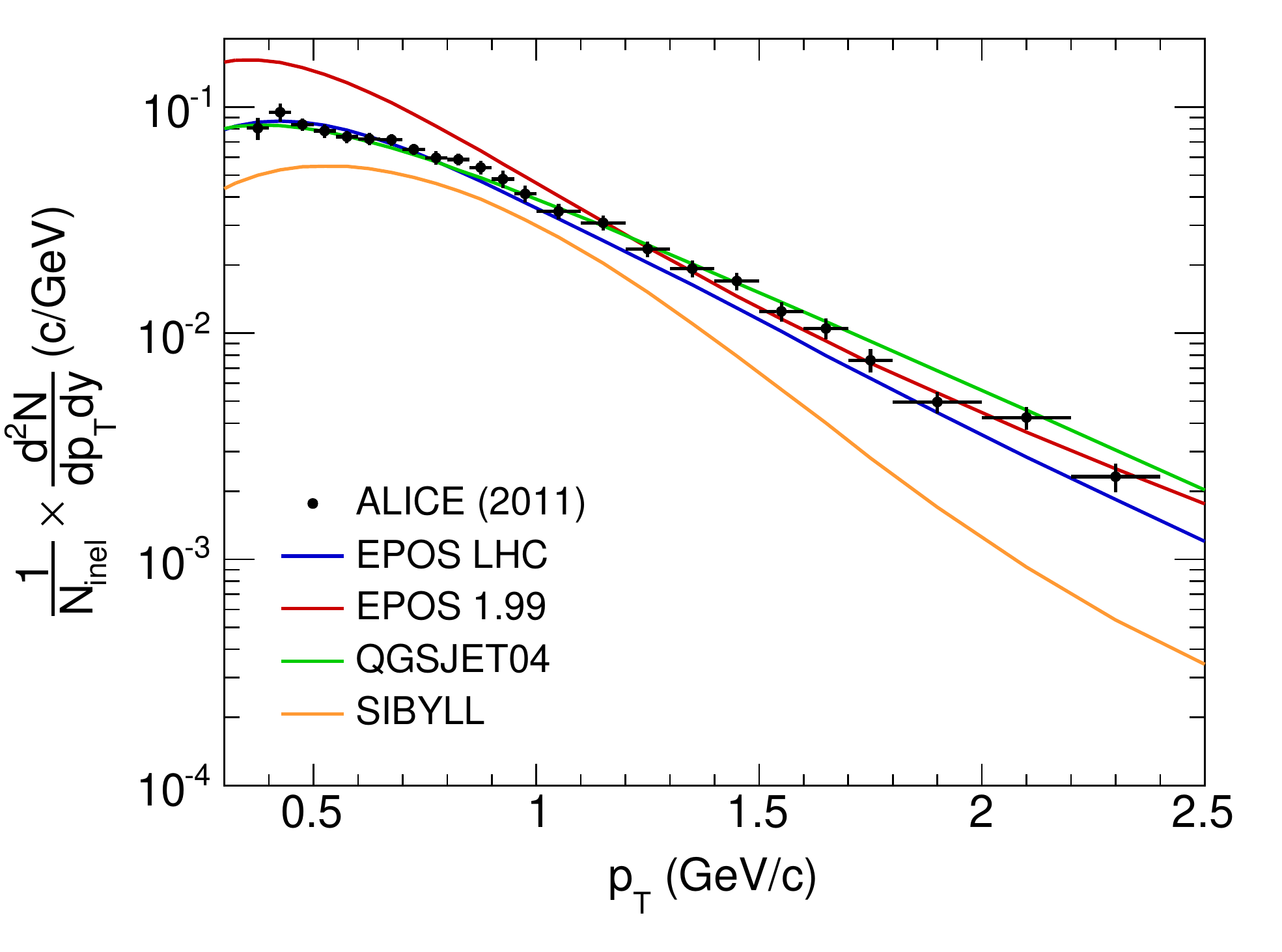}
\caption{
The \pp{}$\rightarrow$\,\pbar{} production yields measured by ALICE\,\cite{Aamodt:2011zj} (dots) 
for rapidities in the range $\left|y\right| < 0.5$ as function of $p_{T}$. 
Corresponding calculation from MC generators are superimposed (lines).  
}
\label{Fig::alice_pp}
\end{figure}

In Fig.\,\ref{Fig::alice_pp} it is compared the \pbar{} production yield measured by ALICE at CERN using \pp{} interactions at LHC with $\sqrt{s} = 900$\,GeV.
Data are shown for the rapidity range $\left|y\right|$ $<$ 0.5 as function of $p_{T}$. 
The comparison with MC generators shows a good agreement with 
\texttt{EPOS LHC} and  \texttt{QGSJET-II-04} at all $p_{T}$-values, and agreement with \texttt{EPOS 1.99} for $p_{T} > 1.2$ GeV/c.

Due to the better agreement with data of \texttt{EPOS} model, \texttt{EPOS LHC} has been chosen as reference model. 
\texttt{EPOS 1.99} is used in the following to cross check the calculation of the systematic error associated to the MC predictions. 

In the laboratory frame the proton momentum for NA49, BRAHMS and ALICE are 
$p_{\textrm{NA49}} = 158$~GeV/c, $p_{\textrm{BRAHMS}} = 21$~TeV/c and $p_{\textrm{ALICE}} = 432$~TeV/c.
The \pbar{} production in the incident proton momentum range between few hundreds of GeV/c and 10 TeV/c 
has not been measured yet, despite this is the range of interest for the 
the estimation of the cosmic rays \pbar{} production in the region accessed to the 
\AMS~\cite{Aguilar:2016pbar} and PAMELA~\cite{PAMELA:Results} \pbarp{} data. 

To evaluate the error associated to the MC prediction due to the discrepancies between the model and the available data we minimize  
separately the $\chi^{2}$ of each experiment NA49, BRAHMS and ALICE with respect to a normalisation factor $k$:
\begin{equation} \label{Eq::chi2_exp}
  \chi^{2} = \sum^{N}_{j} \frac{\left(k f_{j}^{\textrm{MC}} - f_{j}^{\textrm{Data}}\right)^{2}}{\sigma_{j}^{2}}
\end{equation}
where $j$ runs over the $N$ measurements of a single experiment, 
$f_{j}^{\textrm{Data}}$ and $\sigma_{j}$ are the value and error of a single data point,
and $f_{j}^{\textrm{MC}}$ is the corresponding MC prediction for the single data point kinematics.
The estimated $k$ is close to 1 (the discrepancy is in the 2-4\%range) while the normalisation error $\Delta k$ is 4\% for NA49 and about 10\% for the
other two experiments. In the following we will use $\Delta k$ as an estimation of the systematic error associated to data and MC discrepancies. 

To interpolate the error over the entire momentum range, we define a global variable $\chi^{2}_{glb}$ as a function of projectile momentum $p$:
\begin{equation}\label{Eq::chi2_glb}
  \chi^{2}_{glb}\left(p\right) = \sum_{i} w \left(p; p_{i}\right) \chi^{2}_{i}
\end{equation}
where the subscript $i$ runs over the three experiments NA49, BRAHMS and ALICE, and $w_{i}\left(p; p_{i}\right)$ 
is a weighting function that has a gaussian shape with maximum at $p_{i}$ 
and is defined such as $\sum_{i} w_{i}\left(p; p_{i}\right) = 1$ at all momenta $p$. From the $\chi^{2}_{glb}$ distribution $\Delta k$ can be 
estimated for all momenta. 
This procedure has been repeated using the \texttt{EPOS 1.99} model and the two models lead to consistent results. 
%
\begin{figure}[!t]
\includegraphics[width=0.46\textwidth]{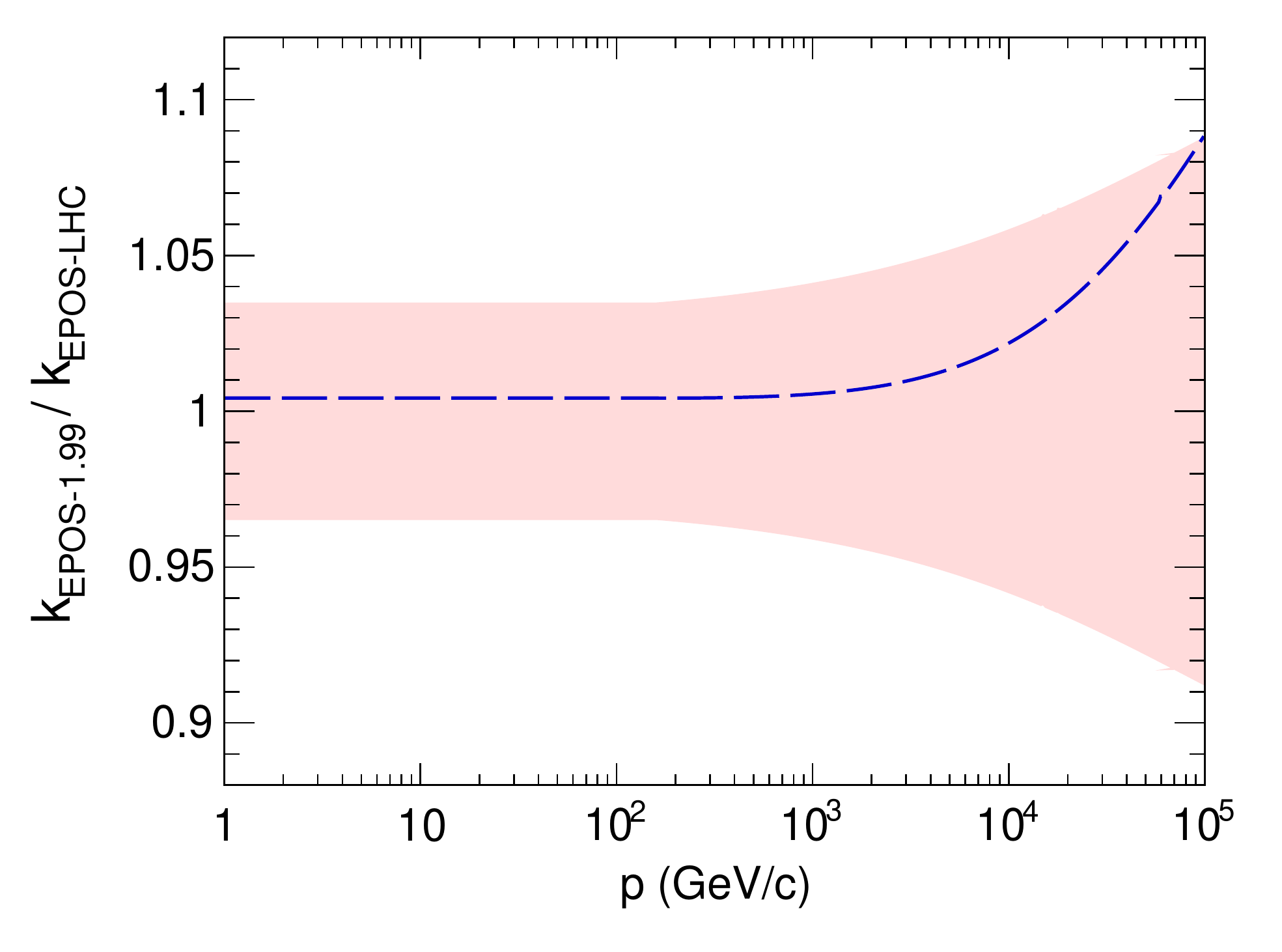} 
\caption{\captionsize%
  Mean ratios between gaussian fits of models and \texttt{EPOS LHC} as a function of projectile
  proton momentum. The shaded area stands for the corresponding sigma of the gaussian.
}
\label{Fig::ccEPOS}
\end{figure}
%
The resulting level of cross-section uncertainty is shown in Fig.\,\ref{Fig::ccEPOS}.
This uncertainty is about 3.5\% at momentum below 160 GeV/c. It increases slowly with momentum to become $\sim$\,9\% at $p=10^5$\,GeV/c.
Up to $10^5$ GeV/c of momentum, the difference between \texttt{EPOS LHC} and \texttt{EPOS 1.99} is within the assigned uncertainties. 

The antineutron yield has also been investigated.
Traditionally it has been assumed that $\sigma_{pp\rightarrow\bar{n}} \equiv \sigma_{pp\rightarrow \bar{p}}$,
which gives an antineutron flux contribution identical to that of direct antiproton production. 
%
\begin{figure}[t] 
\includegraphics[width=0.46\textwidth]{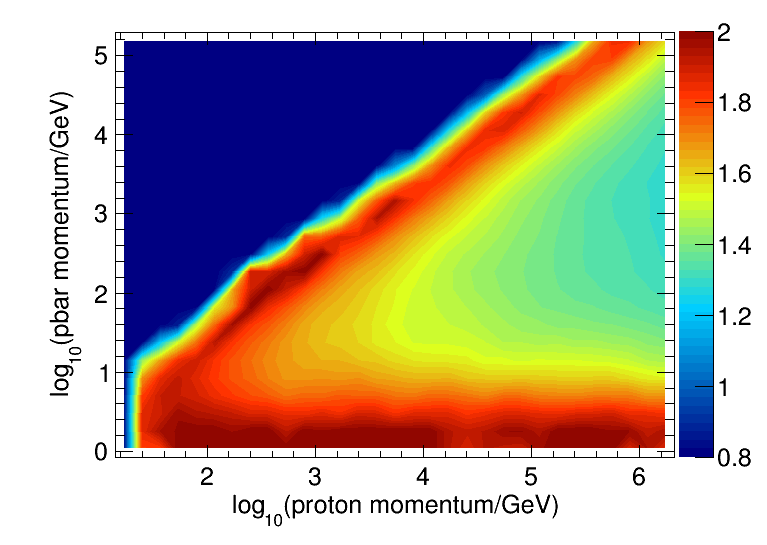} 
\caption{\captionsize%
Mean ratio between antineutrons and antiprotons from \pp{} collisions as function of the 
$p_{\rm pri}$ (primary proton momentum) and $p_{\rm sec}$ (momentum of the secondary antiparticle) predicted by \texttt{EPOS LHC}. 
} 
\label{Fig::ccNbarPbar} 
\end{figure} 
%
Preliminary studies on deuteron-proton interactions have
suggested a possible enhancement of
the antiproton yield from \n-\p{} collisions with respect to \pp{} \citep{Fisher2003},
which would indicate a preferential production of \p-\nbar{} pairs compared to \n-\pbar{} pairs generated in \pp{} collisions.
Based on these data, recent parameterizations proposed a factorized scaling of the type
$\sigma_{pp\rightarrow\bar{n}} \equiv \kappa_{n} \times \sigma_{pp\rightarrow \bar{p}}$,
with $\kappa_{n}= 1.3\pm\,0.2$ \citep{diMauro:2014zea} or $\kappa_{n}=1.37\pm\,0.06$ \citep{Kappl:2014}.   
Under \texttt{EPOS-LHC}, 
the ratio \nbar/\pbar{} is in general not constant over the phase space,
ranging from 1 to 1.9 or more as shown in Fig.\,\ref{Fig::ccNbarPbar}.
However, \texttt{QGSJET-II-04} and \texttt{Sibyll} show no clear preference for \nbar{} production,
similarly to other hadronization models such as PYTHIA or DPMJET \citep{Kappl:2014}.

To account for uncertainties on antineutron production, 
we conservatively assumed a full correlation with the uncertainties in antiproton production.
This contributes appreciably to the errors on the antiproton source term.
Improving measurements on antineutron production is clearly essential for reducing the uncertainties in the CR antiproton flux.

A minor contribution is the one arising from
non-annihilating inelastic collisions of CR antiprotons in the ISM, \ie, 
those processes such as \pbar+\p$\rightarrow$\,X+\pbar$^{\prime}$ where the kinetic energy $E^{\prime}_{\bar{p}}$ of the final state antiproton
is lower than its initial energy $E_{\bar{p}}$. These reactions generate a so-called ``tertiary'' components of CR antiprotons
that we have included using the parameterization of \citet{Tan:1983Ng}.
The energy distribution of tertiary antiprotons is taken proportional to $E^{-1}_{\bar{p}}$, being $E_{\bar{p}}$ the energy 
of the secondary antiprotons. This component is important only in the GeV energy region.

We found that, at high energies, the antiproton source term calculated with \texttt{EPOS LHC} is slightly larger than 
that of \citet{diMauro:2014zea}, 
but the two models are consistent each other within the estimated level of nuclear uncertainties (under the same model of CR transport).
Both models lead to larger antiproton source term in comparison to previous parameterizations.
After accounting for propagation effects, our predicted antiproton flux is larger than that of other works. 
This feature arises from our spatial-dependence modeling of CR diffusion.




\end{document}